\documentclass[12pt,a4paper]{article}
\usepackage[a4paper,margin=1in]{geometry}
\usepackage[T1]{fontenc}
\usepackage[utf8]{inputenc}
\usepackage{tocbibind}
\usepackage{url}
\usepackage{apacite}
\usepackage{graphicx}
\usepackage{float}
\usepackage{subfig}
\usepackage{caption}
\usepackage{makecell}
\usepackage{mathtools}
\usepackage{commath}
\usepackage{booktabs}
\usepackage{multirow}
\usepackage{adjustbox}
\usepackage{enumitem}
\usepackage{placeins}
\usepackage{bbold}
\usepackage{xcolor}

\usepackage[bottom]{footmisc}

\usepackage[doublespacing]{setspace}

\newcommand\blfootnote[1]{%
	\begingroup
	\renewcommand\thefootnote{}\footnote{#1}%
	\addtocounter{footnote}{-1}%
	\endgroup
}

\setlist[enumerate]{label*=\arabic*.}

\makeatletter
\renewcommand\p@subfigure{\thefigure.}
\renewcommand\p@subtable{\thetable.}

\makeatother

\begin{document}

\title{\Large \textbf{ ‘Good job!’ The impact of positive and negative feedback on performance.}}
	\author{ \large
		Daniel Goller$^{1,2}$, Maximilian Späth$^{3}$*\blfootnote{Preliminary version. Do not quote or circulate without permission of one of the authors. Comments are very welcome. We like to thank \textit{Swiss-ski} (Swiss ski federation), Michel Roth, David Morris, Alexander Mesch, and the \textit{DSV} (German swimming federation) for valuable insights into the sports contests from the perspective of (former) professional athletes. Also, we thank participants of the ESEA Conference, 2022, and the Berlin School of Economics Workshop, 2022, as well as Enzo Brox, Lisa Bruttel, and Sandro Heiniger for their helpful comments and suggestions.}
	} 
	\date{\small ~\\ $^1$ Centre for Research in Economics of Education, University of Bern \\ $^2$ Swiss Institute for Empirical Economic Research, University of St.Gallen \\ $^3$ Department of Economics, University of Potsdam \\ ~\\ This version: \today}
	
	\maketitle
	
	\thispagestyle{empty}
	
	\begin{abstract}
		We analyze the causal impact of positive and negative feedback on professional performance. We exploit a unique data source in which quasi-random, naturally occurring variations within subjective ratings serve as positive and negative feedback. The analysis shows that receiving positive feedback has a favorable impact on subsequent performance, while negative feedback does not have an effect. These main results are found in two different environments and for distinct cultural backgrounds, experiences, and gender of the feedback recipients. The findings imply that managers should focus on giving positive motivational feedback. 
	\end{abstract}
	\small
	\vfill
	\newpage

	\newpage
	
	\section{Introduction}\label{sec:Intro}
 
	Providing performance feedback is one of the main tasks of managers and leaders \cite{Morgeson.2010}. One important aim of feedback is to create a favorable emotional response. At best, positive or negative feedback can motivate employees and increase their productivity. In the worst case, it leaves the employees frustrated and unproductive. Therefore, the question of how feedback impacts subsequent performance is of tremendous importance.\par

	Consequently, numerous studies investigating the impact of feedback on creativity \cite{Harrison.2015,Itzchakov.2020,Kim.2020}, the learning process of individuals and firms \cite{Hattie.2007,Lee.2021} or motivation \cite{Deci.1972,Fong.2019} emerged. In particular for positive and negative feedback on performance or productivity, studies show the full range from favorable to unfavorable effects \cite[etc]{Eggers.2019,Kluger.1996,Podsakoff.1989,Sleiman.2020,Waldersee.1994}.\par 
 
    The two major difficulties when investigating the impact of feedback on performance are (1) observing truthful and trustworthy feedback in real-incentive situations and (2) quantifying feedback and performance. While observational studies typically fail to satisfactorily tackle the second difficulty, experimental studies cannot fulfill the first requirement. We are not aware of any causal study in which both requirements are met together.\par 

    To address this common shortcoming, we exploit a unique setting to estimate the causal effect of positive and negative feedback on subsequent performance. For this purpose, we use data from professional sports: diving as the primary data source, and ski jumping for supplementary analyses. In these sports, individuals’ performance is evaluated subjectively by a jury of seven (or five) experienced judges according to precise rules. Each judge independently issues one rating for the task performance (hereafter, "judges rating" or “rating”). Discarding the highest and lowest rating(s), the common assessment of the jury is calculated from the average of the three remaining ratings (hereafter, “jury performance assessment”).\footnote{Receiving the jury performance assessment can already be seen as a \textit{knowledge of results} \cite{Kluger.1996} intervention. The analysis of this knowledge of results, however, is beyond the scope of this paper.} \par 

    Following the definition in \citeA{Kluger.1996}, stating that feedback is information about one’s task performance provided by an external agent, we consider the deviation of the discarded (highest and lowest) ratings from the jury's performance assessment as feedback on task performance. The discarded ratings are not relevant to the assessment of task performance, but this additional information about judges' general perceptions of performance provides feedback that can only work through the motivational channel on subsequent performance. \citeA{Kluger.1996} argue that the feedback sign depends on the relation between the performance rating and a benchmark. In line with this, discarded ratings define quasi-randomly occurring positive (negative) deviations from the jury performance evaluation that serve as positive (negative) feedback. No deviation from the benchmark implies neutral feedback. We describe the evaluation and feedback process in more detail in Section \ref{sec:Var}, Figure \ref{fig:feed_desc}. \par
 
    We test several of the propositions from the model of the seminal work by \citeA{Kluger.1996} within a single framework. In our setup, the feedback is truthful, accurately observable, and from an external source. Feedback can impact subsequent performance only through its motivational impact. Performance is strongly incentivized and can be precisely quantified. The performance is measured in non-artificial tasks that individuals are not only familiar with but that are routine aspects of their work. What is particularly valuable from a management perspective is that we can investigate the impact of feedback in an international context. \par

    Theoretically guided by the feedback intervention model \cite{Kluger.1996}, we investigate the effect of positive and negative feedback on performance. Further, we investigate the internal and external generalizability of the results. To assess internal generalizability, we can use our extensive data to analyze whether situational (or personal) variables and task characteristics moderate the effects of the feedback intervention on performance. The international sample covering female and male individuals from more than 50 nations from 6 continents offer the unique opportunity to analyze feedback effects for different cultural backgrounds and gender within the same framework. To investigate external generalizability, we complement the main findings with a second, independent setting. We investigate these aspects using both classical statistical and causal machine learning methods. This is followed by analyses examining the feedback interventions' long-term, repetition, and spill-over effects.\par
	
	Our analysis shows a performance-enhancing causal effect of positive feedback. The favorable effect of positive feedback is found for recipients from different cultural backgrounds, experience levels, and gender. We observe favorable effects even when individuals repeatedly receive positive feedback. The impact of positive feedback is stronger when the relevance of the task is high. In contrast to all this, negative feedback on average does not have an impact on performance. Merely, the subgroup of the more experienced individuals benefits from negative feedback.\par
 
	Our findings imply that managers can use positive feedback to enhance the performance of their employees. Importantly, positive feedback can be given repeatedly on a regular basis. It has a favorable impact irrespective of several relevant characteristics of the recipient and can be universally applied in an international context. With our main finding we are in line with the studies conducted by \citeA{Azmat.2010}, \citeA{Bandiera.2015}, \citeA{Choi.2018}, and \citeA{Itzchakov.2020} for positive feedback and the meta-study by \citeA{Fong.2019} for negative feedback. We complement decades of research that provides guidelines on how to optimally give feedback \cite{Balcazar.1985, Alvero.2001, Sleiman.2020}.\par

	\section{Theoretical framing}\label{sec:theory}

    To provide a theoretical foundation for the later empirical analysis, we begin by describing the concept of feedback. Then, we collect relevant empirical research and form predictions based on propositions stated by \citeA{Kluger.1996}.
 
	\subsection{The concept of feedback}\label{sec:definition}
	
	Feedback exists in many forms. \citeA{Kluger.1996} define feedback as "[...] actions taken by (an) external agent (s) to provide information regarding some aspect (s) of one's task performance" (p. 255).
	\citeA{Burgers.2015} distinguish between elaborate and simple feedback. Elaborate feedback typically includes a lengthy explanation, which provides a guide for learning. Simple feedback merely gives information, about whether something was done right or wrong.
	\citeA{Burgers.2015} further distinguish between descriptive, comparative, and evaluative feedback. Descriptive feedback -- sometimes called objective feedback \cite{Johnson.2013} -- merely sums up behavior shown by the agent. Comparative feedback uses the performance of other individuals as a reference. Evaluative feedback provides a judgment of the performance. 
	\citeA{Villeval.2020} distinguishes between a cognitive and a motivational perspective. The cognitive perspective rests on the assumption that individuals have imperfect knowledge about their skills. Here, feedback serves as a signal used in an information-updating process. The motivational perspective focuses on the impact of feedback on intrinsic motivation.\par
 
	Individuals might receive feedback from one agent or several agents. \citeA{Stone.1984} find that receiving feedback from two sources instead of one source increases self-perceived task competence. Related, there is a strand of literature analyzing multi-source feedback \cite{Bailey.2002, Smither.2005}, also called 360 degree feedback \cite{DeNisi.2000}. Finally, feedback can be with direct consequences or inconsequential. Often feedback comes without direct (monetary) consequences. Still, research shows that agents also react to irrelevant information \cite{Abeler.2011, Cason.1998}.	\par
 
	The focus of our paper lies on the impact of simple and evaluative feedback on subsequent performance. The feedback is subjective in the sense that is created by subjective evaluation based on objective guidelines. Our study focuses on the impact of single feedback embedded in a multi-source evaluative process. The feedback has no further consequences besides that it can motivate or demotivate the recipient.
	One important distinction is between positive and negative feedback. We define positive feedback, sometimes called promotion-orientated feedback \cite{Carpentier.2013}, as the expression that the evaluated performance is above a certain reference point. We define negative feedback, sometimes called change-orientated feedback \cite{Carpentier.2013} or corrective feedback \cite{Waldersee.1994}, as the expression that the rated performance is below the reference.\par
  
	\subsection{Review and hypotheses} \label{sec:lit}
 
	In their influential model, \citeA{Kluger.1996} assume that there are no behavioral effects when there is no discrepancy between the rating and the reference. Positive feedback increases effort if the agent has the possibility to set new self-goals. Likewise, negative feedback leads to an increase in effort. Similarly, \citeA{Villeval.2020} argues that positive and negative feedback fosters motivation. On the other hand, positive feedback can lead to a decrease in efforts, when individuals have no possibility to set new goals \cite{Kluger.1996}. Negative feedback can discourage individuals when it threatens the self-perception of their competence \cite{Fong.2019}. \par
	
	Some empirical studies show a favorable impact of positive feedback. \citeA{Choi.2018} find a better performance in a computerized task after purely positive feedback than in a baseline treatment. \citeA{Itzchakov.2020} report better performance in a brainstorming task after positive than after neutral feedback. \citeA{Bandiera.2015} report that positive feedback improves the performance of university students and \citeA{Azmat.2010} that positive relative rank feedback enhances the performance of high school students. Other studies, such as  \citeA{Podsakoff.1989} reporting no impact of positive feedback on performance in an object-listing task, find no influence of positive feedback. \citeA{Waldersee.1994} even report an adverse impact of positive feedback on the performance of employees of fast food restaurants.\par
	
	Empirical work on the effect of negative feedback provides an ambiguous picture. Several studies show a favorable impact of negative feedback. As for positive feedback, \citeA{Choi.2018} find an improved performance after purely negative feedback in comparison to a baseline treatment. \citeA{Azmat.2010} find a favorable effect of negative relative rank feedback.
    \citeA{Itzchakov.2020} report a positive impact of negative feedback on performance in a brainstorming task. \citeA{Podsakoff.1989} report a favorable impact of negative feedback in an object-listing task. \citeA{Waldersee.1994} find a performance-enhancing effect of negative feedback for employees of fast food restaurants. Some research, such as the meta-study by \citeA{Fong.2019}, shows no impact of negative feedback. Other studies show an unfavorable impact. For example, \citeA{Deci.1972} observe that a negative feedback group shows lower motivation to conduct a puzzle task than a control group.\par
	
	A reason for the ambiguity in reaction to negative feedback might be heterogeneity in the way how individuals update their perception after receiving self-relevant information. Some research finds that agents do not fully update their self-perception after negative information, while they  update their self-perception after observing a positive signal \cite{Eil.2011, Kuzmanovic.2015, Mobius.2022, Sharot.2012}. This would imply to find no reaction to negative feedback. Yet, other studies observe a rational updating of beliefs for positive and negative information \cite{Barron.2021} or even an overweighting of negative information \cite{Coutts.2019,Ertac.2011}, leaving this strand of empirical research inconclusive.\par

	We build our hypotheses on the theoretical model by \citeA{Kluger.1996}. We argue that in the domain of professional performance, there is always the possibility to set more ambitious goals. This indicates that positive feedback might have a favorable impact. \\ 
 
	\begin{center}
		\textbf{Hypothesis 1 - Positive Feedback: \\ The performance is better after receiving positive feedback than after receiving neutral feedback.}
	\end{center}

	We follow \citeA{Kluger.1996} and \citeA{Villeval.2020} by assuming that also negative feedback has a performance-enhancing effect. We argue that in the field of professional performance, individuals have a rather stable self-perception of confidence.
	\begin{center}
		\textbf{Hypothesis 2 - Negative Feedback: \\ The performance is better after receiving negative feedback than after receiving neutral feedback.}
	\end{center}

    A vital aspect that most empirical studies usually can barely answer is the question of the generalizability of these hypotheses. Here, it is useful to distinguish between the two superordinate layers of personal and task-specific characteristics by which effects could be moderated (compare \citeA{Fong.2019}, for example).\par
 
    For task characteristics, our hypotheses more readily generalize when individuals' responses to feedback are inherently similar irrespective of the difficulty and importance of the task. Difficult and easy tasks might be perceived differently \cite{Moore.2008}, which can lead to different perceptions of feedback \cite{Pulford.1997} and varying subsequent performance \cite{Vancouver.2004}. \citeA{Kluger.1996} argue that the reaction to feedback is stronger the fewer cognitive resources are needed to perform the task. Likewise, performance might differ depending on the importance of the task \cite{GH.2022}. Here, \citeA{Kluger.1996} argue that the effectiveness of feedback increases the more attention is on the task. Guided by the model predictions of \citeA{Kluger.1996}, we do not expect generalizability across task characteristics. Accordingly, we expect stronger feedback effects on  performance for (relatively) easier tasks needing fewer cognitive resources and more important tasks that require more attention.\par
 
    Within the personal domain, three potential moderators seem highly relevant in modern workplaces: cultural background, gender, and experience of the feedback recipients. The literature acknowledges that despite the high relevance of cultural differences in a globalized world, non-WEIRD (not coming from Western, Educated, Industrialized, Rich, and Democratic countries) individuals are largely underrepresented in behavioral research \cite{Henrich.2010}. For example, authors postulate differences in self-construals \cite{Markus.1991}, in feedback seeking of individuals \cite{SullyDeLuque.2000} and in feedback reaction of firms \cite{Rhee.2020} between collectivistic and individualistic cultures.\par  
    
    \citeA{Bear.2017} postulate and \citeA{Berlin.2016}, respectively, \citeA{Roberts.1994} observe different feedback reactions for women than for men. \citeA{Eggers.2019} find that the reaction of organizations to negative feedback depends on the experience in the business area. 
    \citeA{Kluger.1996} propose differential effects for individuals' behavioral or psychological traits. More relevant from a managerial perspective is if those potentially moderating traits are associated with directly observable characteristics of individuals in a company’s diverse context. We refrain from forming explicit expectations and leave the question of generalizability for different cultural backgrounds, genders, and experience levels exploratory.

	\section{Setting and data}\label{sec:data}
	
	We collect data on international competitions of two competitive sports. In the two sports, namely, ski jumping and diving, athletes compete individually in multi-round competitions. In each round, the athletes' task execution is evaluated by multiple professional judges. \par
	
	Besides the similarities, there are several specifics to each of the sports. 
	In diving, athletes acrobatically jump into the water. We use data on individual performances in three different types of competitions: 1m springboard, 3m springboard, and 10m platform. The scoring consists of two elements. First, each jump is rated by seven judges with respect to the proper execution. Each judge can reward up to 10 style points (in increments of 0.5). The two highest and the two lowest judges' ratings are discarded for the jury performance assessment of the jump, for which the remaining three judges' ratings are summed up. Second, the jury performance assessment is multiplied by the difficulty coefficient, which depends on the complexity of the jump and is assigned to the jump according to the official rules.\footnote{See \url{https://resources.fina.org/fina/document/2021/01/12/916f78f6-2a42-46d6-bea8-e49130211edf/2017-2021_diving_16032018.pdf} for a current version of the rules (last accessed on 01/23/2023).}
	In competitions between women, points are accumulated over five jumps, and in competitions between men, over six jumps. Depending on the contest there are preliminary rounds and/or semi-finals and the final round. \par
 
	In the winter sport of ski jumping, athletes jump on skis after sliding down a ramp. Scoring consists of four components. First, athletes receive points for the length of their jump. Second, there are compensation points for the force and direction of the wind. Third, scoring depends on the length of the ramp (gate points). Fourth, athletes receive up to 20 style points (in increments of 0.5) for the flight and landing of the jump. The (style) ratings are independently rewarded by five judges according to official rules.\footnote{See \url{https://assets.fis-ski.com/image/upload/v1665482445/fis-prod/assets/ICR_Ski_Jumping_2022_marked-up.pdf} for a current version of the rules (last accessed on 01/23/2023).} The worst and the best rating are discarded and the other three are accounted for the athletes' score of the round.
	In a typical competition, 50 athletes start in the first round, of which the 30 best reach the final round. After the final round, both jumps' total scores are added to determine the winner and the succeeding rankings. \par	
	
	\subsection{Data sets}\label{sec:data_desc}

	\begin{table}[H]
		\caption{Descriptive statistics} 
		\label{tab:table_desc_red}
		\begin{tabular}{l r r c r r} 
			\toprule
			& \multicolumn{2}{c}{Diving} & & \multicolumn{2}{c}{Ski jumping} \\
			\cline{2-3} \cline{5-6}  
			& Mean & Std. dev. & & Mean & Std. dev. \\ 
			\midrule
			\textit{Panel A: Treatments}    ~~~~~~~~~~~~~~~~~~~~~~~~ &     &    & &         &           \\
			Positive Feedback (deviation positive)  & 0.426  & (0.286)  & & 0.316   & (0.262)   \\
			Negative Feedback (deviation negative)  & 0.477  & (0.320)  & & 0.357   & (0.290)   \\			
			&  &  & &    &  \\
			\textit{Panel B: Outcomes}    ~~~~~~~~~~~~~~~~~~~~~~~~   &     &    & &         &           \\			
			Score                                   & 68.737 & (14.557) & & 118.647 & (16.204)   \\
			Performance (rem. 3 judges’ ratings)             & 7.119  & (1.189)  & & 17.771  & (0.744)   \\
			Performance (all 5 / 7 judges’ ratings)              & 7.110  & (1.182)  & &  17.765 & (0.741)   \\
			&  &  & &    &  \\			
            \textit{Panel C: Covariates}    ~~~~~~~~~~~~~~~~~~~~~~~~ &     &    & &         &           \\
			Compatriot judge                        & 0.248  &          & & 0.457   &           \\
            Home event                              &  0.099 &          & & 0.127   &           \\
            Experience (Age in years)               & 22.429 & (3.789)  & & 26.836  &  (4.949)  \\
            Female                                  & 0.450  &          & &         &           \\
			Difficulty                              &  3.211 & (0.331)  & &         &           \\
			Distance                                &        &          & & 122.608 & (11.837)  \\
			Prev. Distance                          &        &          & & 123.940 & (11.143)  \\
			Prev. Difficulty                        &  3.166 & (0.317)  & &         &           \\ 
			Prev. Performance                     & 7.270  & (0.958)  & & 17.854  & (0.580)   \\			
			\midrule 
			N  &   & 13075 & &   &  4529  \\
			\bottomrule
			\multicolumn{6}{l}{\footnotesize Notes: Mean and standard deviation (in parentheses; for non-binary variables). rem. = }\\
   			\multicolumn{6}{l}{\footnotesize ~~~~~~~~~ remaining. Some variables were only observed in one of the data sets. Full descriptive}\\
			\multicolumn{6}{l}{\footnotesize ~~~~~~~~~   statistics in Appendix Table \ref{tab:table_desc}.}
			
		\end{tabular}
		
	\end{table}	
 
	The main analysis is conducted using data on official diving competitions from 2013 through 2017. This includes special events such as World Championships and the Summer Olympics. 
	Except for the first jump, each jump constitutes one observation. We exclude observations where the rating points of the current or subsequent jump are at the lower or upper bound.\footnote{To put it more concretely: We remove observations that have received an average score of 9.5 or higher (19.5 in ski jumping), as well as those with an average score of less than 5 (14 in ski jumping). Furthermore, we remove observations with individual scores of 3 or lower (14 in ski jumping), as these are most likely to be crashes. All of these choices are robust to changes, and we show the robustness of the results to data pre-processing in the results section.} Athletes who stop competing during the contest are excluded, e.g., due to injury.\par

	We conduct the analysis based on 13075 observations. The data consists of the jumps performed by 434 athletes from 54 countries in Africa, Asia, Europe, North America, Oceania, and South America. As visible in panel C of Table \ref{tab:table_desc_red}, roughly one-half of the athletes are female and on average 22.4 years old. In 25 percent of the cases, at least one of the judges has the same nationality as the task taker and about 10 percent of observations are at a home event. Difficulty and previous difficulty of the jump are on average around 3.2, and (current and previous) performance are on average around 7.1 to 7.3. \par
	
	For our analysis on ski jumping, we have 4529 observations on events from the 2010/11 through 2016/17 season (based on a collection conducted by \citeA{Krumer.2022}). Each observation refers to a second jump. Athletes who fail to qualify for the second round are excluded. In 13 percent of the cases, athletes perform in their respective country of birth. In 45 percent of the cases, one of the judges is of the same nationality as the performing athlete. The average age is about 26.8 years. Jumps are on average about 123 meters and (current and previous) performance are on average around 17.7 (see panels B and C of Table \ref{tab:table_desc_red}).\par

	\subsection{Variables}\label{sec:Var}

    		\begin{figure}[H]
		\centering
		\caption{Illustration of the evaluation and feedback process}
		\label{fig:feed_desc}
			\centering
			\includegraphics[width=0.98\textwidth]{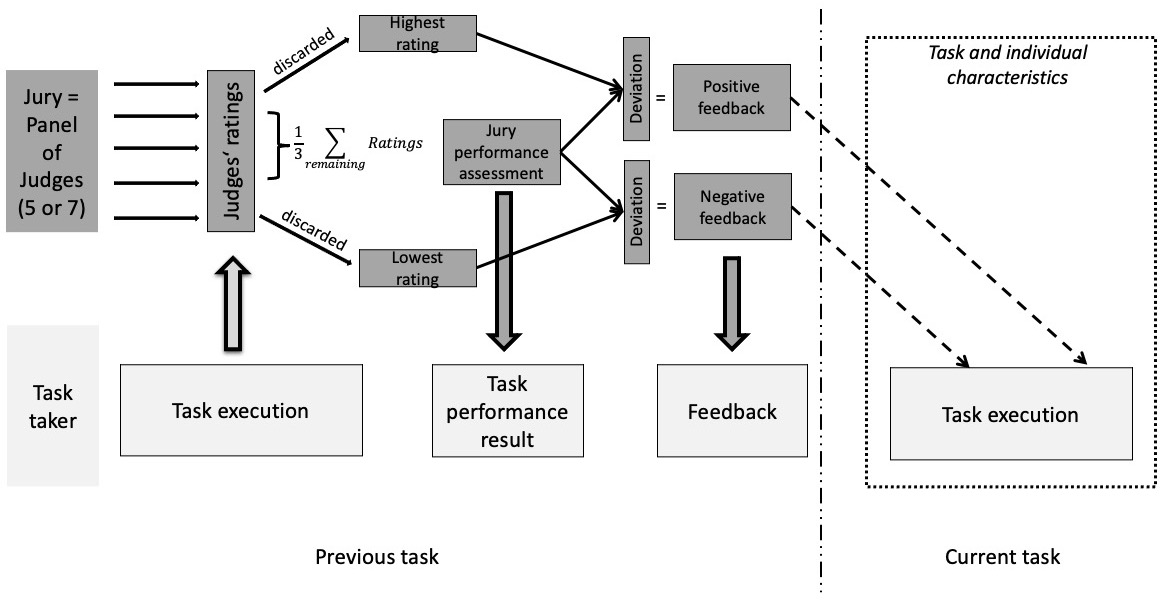}\\ 
		\footnotesize Notes: For a current task (on the right), feedback is given for the previous task (left). The broken ~~~~~~~~  \\
		\footnotesize ~~~~~~ arrows represent our main hypotheses, i.e., the potential influence of feedback on performance\\
  		\footnotesize ~  in the subsequent task. Task and individual characteristics (dotted square) potentially ~~~~~ \\
      	\footnotesize ~~~~~  moderate this effect. In the case of seven judges, the two highest and lowest ratings are ~~~~~~~ \\
        \footnotesize ~~~~~~~ discarded, and only the most extreme ratings are used. See Section \ref{sec:robu} for other specifications \\
		\footnotesize ~  used in the robustness checks. ~~~~~~~~~~~~~~~~~~~~~~~~~~~~~~~~~~~~~~~~~~~~~~~~~~~~~~~~~~~~~~~~~~~~~~~~~~~~~
	\end{figure}

    Figure \ref{fig:feed_desc} describes the evaluation and feedback process in our setup.
    For the task execution evaluation, each judge in the jury independently gives a numerical rating for the task execution of the task taker. The largest and smallest of those judges' ratings are discarded and the jury performance assessment is the mean of the remaining (three) judges' ratings. The task performance assessment quantifies the task performance result. \par 
    
    In our study, we focus on the discarded judges' ratings that are not regarded for the jury's performance assessment and can affect subsequent performance only through their motivational impact. Our treatment variables are constructed as deviations of the discarded judges' ratings from the jury performance assessment. More concrete, \textit{Deviation positive} is constructed by subtracting the jury performance assessment (the mean of the ratings in absence of the discarded ratings) from the largest discarded judges' rating. \textit{Deviation negative} is constructed by subtracting the smallest discarded judges' rating from the jury performance assessment.\footnote{Additionally, we construct and test two alternative specifications. All specifications can be found in the full descriptive statistics in Appendix Table \ref{tab:table_desc}. Especially, for diving, there are two (highest/lowest) judges' ratings discarded. The base specification uses the most extreme judges' ratings. Other specification descriptions and results for the robustness of the alternative treatment variable specifications can be found in Section \ref{sec:robu}.}
    We define \textit{Deviation positive} as positive feedback and \textit{Deviation negative} as negative feedback. Panel A in Table \ref{tab:table_desc_red} provides an overview of the main treatment variables. Both feedback variables, with mean values of 0.426 (0.316) for positive feedback and 0.477 (0.357) for negative feedback, range from 0 (for neutral feedback) to 2.5 (for increasingly positive/negative feedback).\par
    
    To measure the effect of feedback on subsequent task execution, we use the jury's performance assessment that the task takers receive for their subsequent performance (hereafter, "Performance") as our outcome variable. An alternative variable to measure subsequent performance is the mean of the ratings from all (5 or 7) judges.

	\section{Empirical strategy}\label{sec:es}
	
	We study how positive and negative feedback affect subsequent performance. To this end, our identification strategy relies on conditional idiosyncratic variations in the differences between the jury performance assessment and the discarded ratings. This positive (negative) deviation is irrelevant to the assessment of the task performance but provides feedback in the form of additional information about the judges' general perception of the performance.\par

    The identification strategy presumes that, once we condition on a few observable characteristics, there are no omitted influences that are correlated with both outcome, i.e., performance in the task, and treatment, i.e., the positive/negative deviation (feedback for the previous task). Our approach formalizes to the following linear baseline model:
    \begin{equation*}
        Y_i = \alpha + \beta_{+} A^+_i + \beta_{-} A^-_i + \gamma X_i + \epsilon_i,
    \end{equation*}
    
    where the outcome, $Y_i$, is the performance in the (current) task for individual $i$. The continuous treatments $A^{+/-}_i$ are defined as the positive/negative feedback for the (previous) task, and $\beta_{+/-}$ are the coefficients of interest to investigate our hypotheses 1 and 2. $X_i$ contains (pre-determined) covariates of individual $i$ that we need to control for. $\epsilon_i$ is an idiosyncratic error term.\par 
    
    To give credence to the unconfoundedness assumption, we address concerns raised in the literature about potential biases in subjective ratings. First, we consider nationality bias \cite{Heiniger.2021,Krumer.2022,Sandberg.2018,Zitzewitz.2006}, i.e., a judge from the same country as the task taker rates the compatriot better than other individuals. To account for potentially more positive ratings from judges who are compatriots, we include a) a binary variable indicating whether a judge on the panel is a compatriot of the task taker, and b) an indicator if the individual competes in a home event in $X_i$.\footnote{Judges' decisions regarding possible bias in favor of compatriots might be different in front of a supportive crowd \cite{Page.2010,Goller.2020}.} To alleviate remaining concerns about bias based on common nationality, we conduct two further checks. A balancing test in Table \ref{tab:table_bala_test} shows no balancing issues related to compatriot judges. To ensure that the results are not driven by individuals that are potentially subject to nationality bias, we perform a robustness check in which the affected task takers are removed from the sample.\footnote{The results for this can be found in Table \ref{tab:table_robu} and hardly differ materially from the main results.} \par

	Second, there is evidence in the literature of an order of action bias \cite{Damisch.2006,Ginsburgh.2003}. Subjective ratings are found to be affected by the order of task performance, which threatens our identification when some but not all judges are affected. We account for this by controlling for the order in which individuals perform tasks (starting order).
    Third, more difficult tasks were found to be rewarded with higher scores--the difficulty bias \cite{Morgan.2014}. The difficulty of a task in our case is precisely measurable and predetermined. Specifically, in diving, we control for the difficulty of the jump (chosen a priori); in ski jumping, we control for the (previous and current) wind and gate, i.e., the length of the hill--both factors that can influence difficulty and subjective evaluation.\par
	
    Fourth, there could be reputation bias \cite{Findlay.2004}. This bias can lead to better ratings for well-established individuals who typically have a better reputation. To ensure conditional independence, we take into account a) individual and individual-by-season fixed effects and b) current rank in the competition.
    Fifth, the accuracy of subjective performance ratings is found to vary for different performance qualities \cite{Heiniger.2021}. Therefore, we include the individual mean and standard deviation of the jury's performance assessment of the previous task in $X_i$. \par
	
	While not testable, we are confident that the conditional independence assumption is satisfied. Still, we offer two types of checks for it.
	First, in a total of 20 balancing checks in Table \ref{tab:table_bala_test}, only one statistically significant test indicates a solid balancing among observable characteristics. 
	Second, with respect to unobservable characteristics, we provide an indirect approach to support the conditional independence assumption by implementing a placebo treatment test. We replace the treatment variable with a pseudo-treatment variable recorded in the future. The task performance cannot be influenced by the feedback given in the future of this task. Therefore, if we observe all confounding influences, the placebo treatment effect should be zero. If we reject this placebo null hypothesis this points to some unobserved confounding (or other issues like endogeneity or reverse causality), while not rejecting gives some evidence that the conditional independence assumption is plausible.
	Table \ref{tab:table_placebo_div} shows that this placebo test cannot reject our assumption of unconfoundedness. \par
	
	To estimate the main effects of interest, we use linear regression and cluster standard errors on the individual level. In the second step, we apply a method from the causal machine learning literature. 
	For this research, the importance of investigating potential non-linearities in the effect lies in the differently observed treatment intensities, i.e., high or low quantified feedback, for which it is unclear if an estimated constant treatment effect reflects various treatment intensities properly.\par 
	
	With the non-parametric kernel method for continuous treatment effects introduced by \citeA{Kennedy.2017} we investigate the effects for different intensities of the treatment. The method builds on two steps. First, a (doubly-robust) pseudo-outcome is constructed as follows:
    \begin{equation*}
        \xi(\pi,\mu) = \frac{Y-\mu(X,A)}{\pi(A|X)} \int \pi(A|x)dP(x) + \int \mu(x,A)dP(x),
    \end{equation*}	
	where the nuisance functions $\pi(A|X)$ and $\mu(X,A)$ are estimated using a random forest estimator \cite{Breiman.2001}. The pseudo-outcome $\xi(\pi,\mu)$ is doubly-robust in the sense that only (at least) one of the two nuisances needs to be consistent, not both, and is free from confounding influences. In the second step, the average potential outcome for given treatment levels is estimated using a non-parametric kernel regression of the pseudo-outcome on the continuous treatment variable: $E(Y^a)=E(\xi(\pi,\mu|A=a))$.

	\section{Results}
	\label{sec:results}

	\subsection{Main results}\label{sec:results_main}
	
	Our first main finding is that positive feedback is enhancing (subsequent) performance. Panel A in Table \ref{tab:table_base} shows a statistically significant and positive coefficient for positive feedback. The effect is robust to the inclusion of different sets of covariates. In each specification, the average effects are statistically significant at the 1\% level. Panel B replicates this finding for our second data set. As our second main finding, we observe that negative feedback causes an effect close to zero in both panels and all specifications. We do not see any effect of negative feedback on performance.\par

	\begin{table}[H]
		\centering
		\caption{The effect of feedback on performance -- sensitivity to different specifications} 
		\label{tab:table_base}
		\begin{tabular}{l r r r r r r r} 
			\toprule

			\textbf{Performance}   &  \multicolumn{1}{c}{(1)}   & ~~~~  &  \multicolumn{1}{c}{(2)}    & ~~~~ & \multicolumn{1}{c}{(3)}  & ~~~~  & \multicolumn{1}{c}{(4)} \\ 
			\cline{2-2} \cline{4-4} \cline{6-6} \cline{8-8}\\
			\multicolumn{8}{l}{\textit{Panel A: Diving (N=13075)}}  \\ 
			Positive Feedback ~~~~ & 0.242*** & ~~&  0.208*** & ~~& 0.115*** & ~~& 0.100***  \\
			                    & (0.036)   & ~~&  (0.034)  & ~~& (0.032)   & ~~& (0.035)   \\
			Negative Feedback   &  0.018    & ~~&   0.024   & ~~& 0.001    & ~~& 0.007     \\
			                    & (0.030)   & ~~&  (0.030)  & ~~& (0.029)   & ~~& (0.030)   \\
			\rule{0pt}{0.2ex}\\
			\multicolumn{8}{l}{\textit{Panel B: Ski jumping (N=4529)}}  \\ 
			Positive Feedback   & 0.201*** & ~~& 0.180*** & ~~& 0.145*** & ~~& 0.107*** \\
			                    & (0.035)  & ~~&  (0.036) & ~~& (0.034)  & ~~& (0.034)  \\
			Negative Feedback   & -0.063   & ~~&  -0.055  & ~~& -0.049   & ~~& -0.026  \\
			                    & (0.043)  & ~~&  (0.041) & ~~& (0.037)  & ~~& (0.041)  \\ 
		    \rule{0pt}{0.2ex}\\ 
			\midrule 
			Base Covariates         & x & & x & & x & & x     \\
			All Covariates          &   & & x & & x & & x     \\
			Individual Fixed Effect    &   & &   & & x & &       \\ 
			Individual x Season FE     &   & &   & &   & & x     \\ 
			
			\bottomrule
			\multicolumn{8}{l}{\footnotesize Notes: Linear regression. Full regressions in Tables \ref{tab:table_base_ski} and \ref{tab:table_base_div}. All regressions contain previous'}\\
			\multicolumn{8}{l}{\footnotesize  ~~~~~~~~~   jumps jury assessment (\textit{Base Covariates}). \textit{All Covariates} include prev. jumps wind and }\\
			\multicolumn{8}{l}{\footnotesize ~~~~~~~~~  gate points and distance (ski jumping) or difficulty (diving). Also, points behind,   }\\
			\multicolumn{8}{l}{\footnotesize ~~~~~~~~~   compatriot judge, home event, current ranking, SD of previous performance, and start  }\\
			\multicolumn{8}{l}{\footnotesize  ~~~~~~~~~ order. Standard errors are  clustered on the individual level. *, **, and *** represents   }\\
			\multicolumn{8}{l}{\footnotesize ~~~~~~~~~ statistical significance at the 10 \%, 5 \%, and  1 \% level, respectively.}
			
		\end{tabular}
	\end{table}

	The performance-enhancing impact of positive feedback is rather insensitive to the inclusion of more covariates and fixed effects. We start with controlling only for performance in the previous task in column (1). In column (2) we add several control variables as discussed in Section \ref{sec:es}.
	Columns (3) and (4) add individual fixed effects and individual-by-season fixed effects to the regressions. Detailed result tables can be found in the appendix in Tables  \ref{tab:table_base_ski} and \ref{tab:table_base_div}, and for the sake of simplicity, all of the following regressions are based on the specification used in column (3).\par

	\begin{figure}[H]
		\centering
		\caption{Non-linear estimation of feedback on performance}
		\label{fig:nl_main_res}
		\begin{minipage}{0.47\textwidth}
			\centering
			\includegraphics[width=0.95\textwidth]{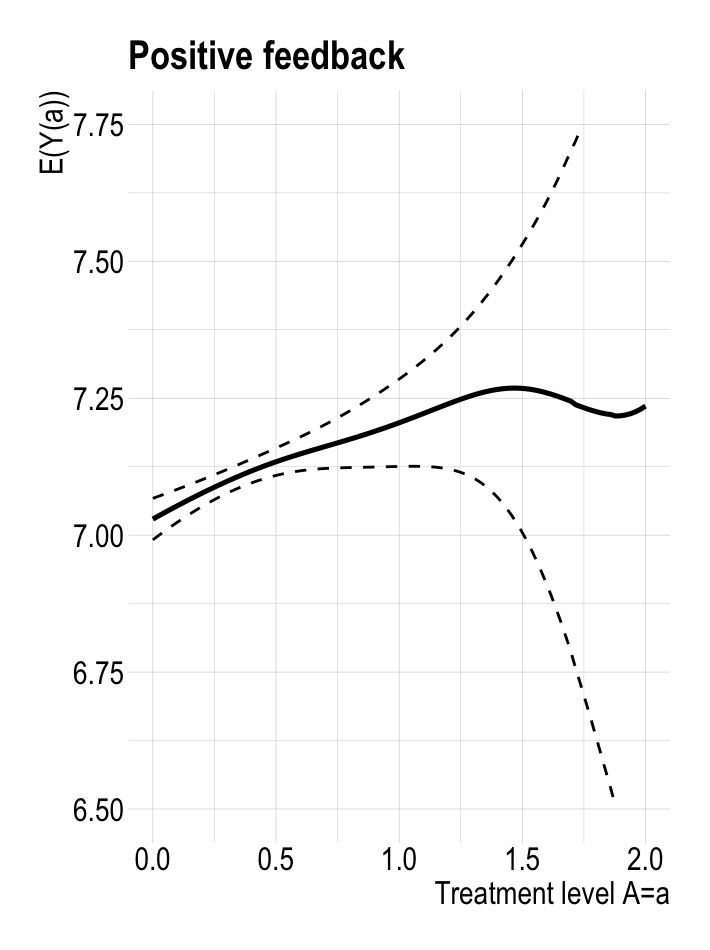} 
		\end{minipage}\hfill
		\begin{minipage}{0.47\textwidth}
			\centering
			\includegraphics[width=0.95\textwidth]{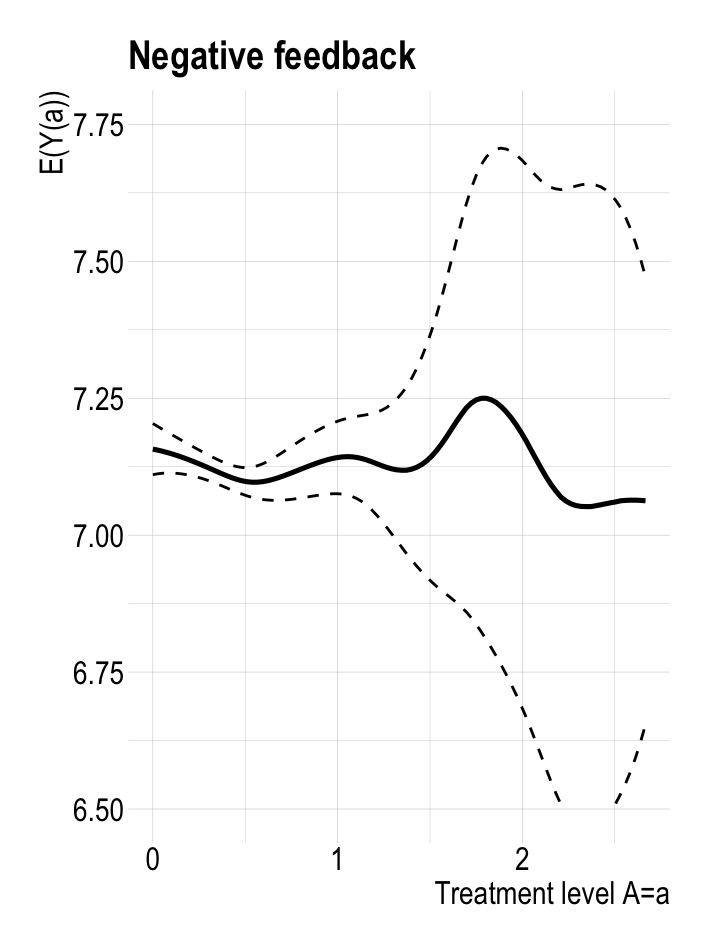}
		\end{minipage}
		\footnotesize Notes: Non-parametric kernel regression for different levels of positive (left) and negative (right). \\
		\footnotesize ~~~~~~~ feedback. Expected outcomes (y-axis) and treatment levels (x-axis) are displayed. ~~~~~~~~ \\
		\footnotesize ~~~~~~~~~~~~ Kernel bandwidths are 0.300 (left) and 0.214 (right) and are determined in a data-driven ~~~~ \\
		\footnotesize ~~~ approach using a cross-validation method. To obtain treatment effects, one might ~~~~ \\
		\footnotesize ~~~~~   calculate the difference of the expected outcomes for two  treatment levels and divide ~~ \\
		\footnotesize ~~~  this by the difference in the treatment levels (treatment intensity). Diving data.  ~~~~~ \\
		\footnotesize ~~~~~~~~~~~~~~  The broken lines represent the 90\% confidence intervals. ~~~~~~~~~~~~~~~~~~~~~~~~~~~~~~~~~~~~~~~~~~~~~~~~
	\end{figure}
	
	Our results show that, on average, positive feedback is enhancing performance. In the following, we go beyond average effects and investigate the effect of positive and negative feedback for different magnitudes of feedback. Figure \ref{fig:nl_main_res} provides non-linear estimates of positive and negative feedback showing the expected outcome (performance) against the extent of the feedback, i.e., the level of the treatment. The (treatment) effect of different feedback intensities can be calculated as the difference in expected outcomes for an increase from some treatment level to another.\footnote{For two different treatment levels $A=a_1$ and $A=a_0$, the effect can be calculated as $\theta(a_1,a_0) = \frac{E(Y(A=a_1))-E(Y(A=a_0))}{a_1-a_0}$. The treatment intensity in this example is $a_1-a_0$, while for a complete picture, it needs to be clear that the treatment level from which the treatment intensity is evaluated is $a_0$ here.}
	In the graph on the left, the effect of positive feedback is positive throughout all feedback intensities, i.e., the expected outcome increases almost steadily as the level of treatment increases.
	With negative feedback, on the right side of Figure \ref{fig:nl_main_res}, the effect varies slightly up and down for different treatment intensities -- although the effect does not appear to be different from zero for any treatment intensity, consistent with the average effect of zero reported in Table \ref{tab:table_base}.
	For both estimations, we find that the linearity assumption in the regression analyses is a good approximation for the non-linear effect curves.
	Still, especially for the higher treatment intensities the confidence intervals become large and conclusions become imprecise--a fact to which global linear regression models do not give any hint.\par

	Overall, the results provide support for hypothesis 1: The performance is better after receiving positive feedback than after receiving neutral feedback. Contrarily, we do not find support for hypothesis 2, i.e., the performance is not better after receiving negative feedback than after receiving neutral feedback. In the next section, we test if the positive effect of positive feedback and the null effect of negative feedback persists in different sub-populations and is generalizable for diverse personal or situational conditions.

	\subsection{Sub-population and context heterogeneity}

    In the feedback-intervention model of \citeA{Kluger.1996}, as well as, for example, in the meta-study of \citeA{Fong.2019} aspects are collected for which the effects of feedback potentially differ. Personal characteristics, situational aspects, and task characteristics, among other factors, might shape the reaction of individuals to positive and negative feedback. A strength of our unique data set is that it allows us to investigate if we can generalize the results of our analysis. \par
    
    Panel A of Table \ref{tab:table_subpop} exhibits that positive feedback has a favorable impact irrespective of individuals' personal characteristics. We consider three categorizations of the individuals' cultural backgrounds. First, we report that the favorable effect of feedback on performance is present for individuals from WEIRD and non-WEIRD countries.
    Second, we find a favorable impact of positive feedback irrespective of the relative cultural distance to the U.S..
    Third, individuals coming from relatively individualistic and relatively collectivistic countries both react favorably to positive feedback.\footnote{We classify (non-)WEIRD countries according our own assessment based on \citeA{Henrich.2010}; the respective list can be obtained upon request. For cultural distance to the U.S., we use the metrics provided in Table 1 in the research article by \citeA{Muthukrishna.2020}. For individualistic and collectivistic countries, we use data from the index created by \citeA{Hofstede.2011}.} Other personal characteristics that we investigate are experience and gender. We find a performance-enhancing effect of positive feedback for both the relatively more and less experienced. Similar to \citeA{Bear.2017}, we also explore whether there are gender differences in the reaction to feedback. We find that both sexes react favorably to positive feedback For none of the three different definitions of cultural background, nor gender and experience, do the two-sample WALD tests show statistically significant differences. This leads to the conclusion that the effects of feedback are consistent and generalizable across these three personal characteristics.\par
    
    Importantly, we find some heterogeneity with respect to the characteristics of the task. Contested situations offer greater incentives to perform \cite{GH.2022}, with higher task focus and more pressure. Panel B of Table \ref{tab:table_subpop} shows large and positive effects for positive feedback in close competitions, but an insignificant effect for situations that are less competitive. This is in line with the argumentation by \citeA{Kluger.1996} and our expectations. Contrary, we find no support for differential effects for the difficulty of the task. Positive feedback leads to a performance-enhancing impact for easy and hard tasks.\par
    
    The results of the heterogeneity analysis on the impact of negative feedback are largely in line with the main finding. The second column of Table \ref{tab:table_subpop} shows a null effect of negative feedback for most subgroups and all contexts. The only exception is the experience of the individuals, where we find that relatively more experienced individuals improve their performance after receiving negative feedback. A two-sample Wald test (in square brackets) shows that the difference in the reaction between the more and less experienced individuals is statistically significant. The favorable impact of negative ratings for experienced individuals is in line with findings by \citeA{Eggers.2019} on the firm level.

	\begin{table}[H]
		\begin{center}
			\caption{Differential effects}
			\label{tab:table_subpop}
			\begin{tabular}{l r c r} 
				\toprule
				& Positive Feedback & ~~~~~ & Negative Feedback \\
				\cline{2-2} \cline{4-4} 
				\rule{0pt}{0.2ex} \\
                \multicolumn{4}{l}{\textit{Panel A: Individuals` characteristics}} \\
                \rule{0pt}{0.2ex}\\
				WEIRD$^1$ (N=4955)      &   0.086*   (0.048)    & & 0.006 (0.049)  \\
				Non--WEIRD (N=8120)    &   0.135*** (0.043)    & & 0.004 (0.037)  \\
				                        &   [0.447]             & & [0.974]         \\
				\rule{0pt}{0.2ex}\\
				Culturally close to U.S.$^2$ (N=6223)    &   0.132*** (0.046)    & & -0.007 (0.047)  \\
				Not culturally close to U.S. (N=6852)    &   0.101**   (0.044)    & &  0.008 (0.037)  \\
				                        &   [0.626]             & & [0.802]         \\
				\rule{0pt}{0.2ex}\\
				Individualistic country$^3$ (N=6013)    &   0.096** (0.047)    & & 0.007 (0.045)  \\
				Collectivistic country (N=6872)    &   0.144***   (0.045)    & &  0.001 (0.040)  \\
				                        &   [0.461]             & & [0.921]         \\
				\rule{0pt}{0.2ex}\\
				More experienced (age $\geq$ 23y, N=6176) &   0.146*** (0.045)    & & 0.076* (0.039)  \\
				Less experienced (age $<$ 23y; N=6899) &   0.081*   (0.047)    & & -0.062 (0.044)  \\
				                        &   [0.318]             & & [0.019]         \\
				\rule{0pt}{0.2ex}\\ 
				Female (N=5885)         &   0.087*   (0.047)    & & -0.028 (0.042)  \\
				Male (N=7190)           &   0.128*** (0.043)    & &  0.018 (0.039)  \\
				                        &   [0.520]             & & [0.422]         \\	
				\rule{0pt}{0.5ex}\\
                \multicolumn{4}{l}{\textit{Panel B: Task characteristics}}\\
                \rule{0pt}{0.2ex}\\
				Tight competition$^4$ (N=5118)  &  0.173***  (0.056)    & & -0.033 (0.052)  \\
				Non--tight competition (N=7957) & 0.064 (0.039)   & &  0.007 (0.037)  \\
				                        &  [0.110]              & & [0.531]         \\
                \rule{0pt}{0.2ex}\\
				Easy task$^5$ (N=7267)  &   0.154*** (0.043)    & & -0.027 (0.037)  \\
				Hard task (N=5808)      &   0.086* (0.048)      & & 0.025 (0.044)  \\
				                        &   [0.291]             & & [0.366]         \\
				\bottomrule
				\multicolumn{4}{l}{\footnotesize Notes: Linear Regression estimates. Diving data. Control variables as in column (3) in Table \ref{tab:table_base}.}\\
				\multicolumn{4}{l}{\footnotesize ~~~~~~~~~   Standard errors are clustered on the individual level. *, **, and *** represents statistical}\\
				\multicolumn{4}{l}{\footnotesize ~~~~~~~~~   significance at the 10 \%, 5 \%, and 1 \% level, respectively. P-value of WALD test for  }\\
				\multicolumn{4}{l}{\footnotesize ~~~~~~~~~  equality in square brackets. $^1$Western, Educated, Industrialized, Rich, Democratic. $^2$Cultural}\\
				\multicolumn{4}{l}{\footnotesize ~~~~~~~~~  closeness is divided at the median level of an index taken \citeA{Muthukrishna.2020}. $^3$Divided}\\
				\multicolumn{4}{l}{\footnotesize ~~~~~~~~~  at median level of an individualism index constructed by \citeA{Hofstede.2011}; (some countries }\\
				\multicolumn{4}{l}{\footnotesize ~~~~~~~~~  missing). $^4$Athlete is within ten points to first place in final, and to the cut-off in preliminary }\\
				\multicolumn{4}{l}{\footnotesize ~~~~~~~~~  rounds. $^5$Easy and hard according to the median chosen difficulty of the (assessed) task.}
			\end{tabular}
		\end{center}
	\end{table}

	\subsection{Repetition and long-term effects}
	
	For practitioners, it is crucial to know about the impact of feedback when it is given repeatedly and about its long-term effect. Fortunately, our data allows for analyzing the impact of feedback on performance in a repeated setup.\par
 
\begin{figure}[H]
		\centering
		\caption{A non-diminishing effect of positive feedback}
		\label{fig:effect_past_posneg}
			\centering
			\includegraphics[width=0.58\textwidth]{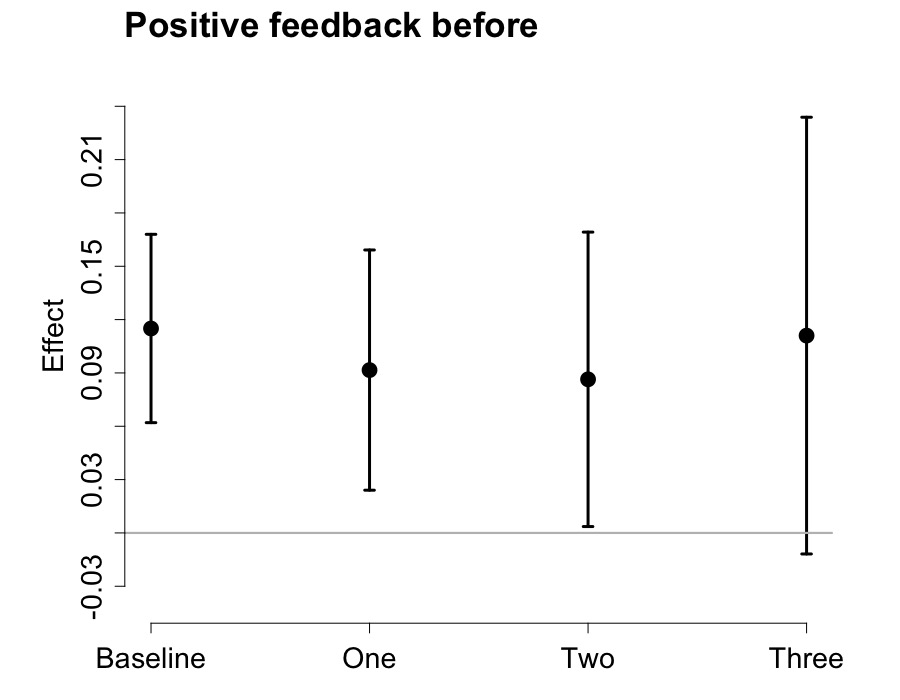}\\ 
		\footnotesize Notes: Linear regression estimates. Diving data. Specifications as in column (3) in Table \ref{tab:table_base}. ~~~  \\
		\footnotesize ~~~~~~~ Standard errors are clustered on the individual level. Effect among those that  ~~~~~~~~~~  \\
		\footnotesize  ~  experienced positive feedback at least one, two, or three times (in the respective  \\
		\footnotesize  ~~~~~~  round) before. The whiskers mark the 90 \% confidence intervals. ~~~~~~~~~~~~~~~~~~~~~~~~
	\end{figure}

	Figure \ref{fig:effect_past_posneg} shows that the favorable impact of positive feedback is non-diminishing with repetition. As a benchmark, \textit{Baseline} shows the average effect of receiving feedback as reported in Table \ref{tab:table_base}, which is not conditional on further previously received feedback.
    We find that for those who have received positive feedback at least one time before, further positive feedback continues to have a positive impact on their performance.
    Similarly, we find a positive influence of positive feedback if the individual has received positive feedback at least two or three times before.

	Table \ref{tab:table_lag_diving} shows the non-persistence of the effect of positive feedback on performance. For reference, column (1) reports the baseline effect for the performance in the task that is conducted directly after the feedback is received. Columns (2-4) provide estimates for the effect of feedback on performance in tasks carried out thereafter. For all follow-up tasks, we find statistically insignificant effects. This indicates that the favorable short-term effect of positive feedback does not carry on to future tasks. Negative feedback has no impact, neither on subsequent nor future tasks. \par

		\begin{table}[H]
		\centering
		\caption{A non-persistent effect of feedback on performance} 
		\label{tab:table_lag_diving}
		\begin{tabular}{l r r r r r r r} 
			\toprule
			\textbf{Performance} ~~~~~~~~ & (1)   & ~~~~~ & (2)  & ~~~~~ & (3)   & ~~~~~ & (4)     \\
			\midrule 
			Positive feedback   & 0.115***  & & -0.010  & & 0.073   & & -0.062 \\
			                     & (0.032)   & & (0.047) & & (0.050) & & (0.061) \\
			Negative feedback   & 0.001    & &  -0.049 & & 0.023   & & -0.079     \\
			                     & (0.029)   & & (0.036) & & (0.046) & & (0.056)   \\ 
			
			\midrule
			Periods after feedback:  & 1         & & 2     & & 3       & & 4     \\
			\midrule
			
			N                        & 13075      & & 10130    & & 7350      & & 4512 \\ 
			\bottomrule
			\multicolumn{8}{l}{\footnotesize Notes: Linear Regression on future outcomes. Diving data. Specifications as in column (3) in  }\\
			\multicolumn{8}{l}{\footnotesize ~~~~~~~~~ Table \ref{tab:table_base}. Standard errors are clustered on the individual level. *, **, and *** represents  }\\
			\multicolumn{8}{l}{\footnotesize ~~~~~~~~~  statistical significance at the 10 \%, 5 \%, and 1 \% level, respectively. }
		\end{tabular}
	\end{table}

	\subsection{Spillover effects on related tasks}
	
    Previously presented evidence shows the favorable effects of positive feedback on the task for which the feedback was obtained. In practice, individuals might do several tasks simultaneously, or a task containing different elements, that potentially influence each other. 
    For example, \citeA{HEcht.2012} show spillover effects of incentive schemes in one task on a related, simultaneously conducted second task. 
    Our settings allow us to study, both, a single-task and a multi-task environment.  \par
    
    Panel A presents the results for the single-task setup. As presented previously in Table \ref{tab:table_base}, we find a performance-enhancing impact of positive feedback and no impact of negative feedback on performance. The difficulty is fixed ex-ante. That we find no impact of feedback on the difficulty can be regarded as a placebo outcome test and supports our identification strategy. Difficulty and performance evaluation jointly determine the combined outcome. Consequently, we observe a favorable effect of positive feedback on the total score.\par
    
    Panel B exhibits the results for the multi-task environment. We observe favorable spillover effects. Receiving positive feedback in Task 1 enhances subsequent performance in Task 1 and the related Task 2. Negative feedback has no impact on either of the tasks.
    In the setup, performance in Task 1 and Task 2 are the most important determinants of combined success and the only ones that can be influenced by the task taker. Consistently, we also find a favorable influence of positive feedback on the total score.\par

		\begin{table}[H]
		\caption{Spillover effects} 
		\label{tab:table_spill}
		\begin{tabular}{l c c c c c} 
			\toprule
			\multicolumn{4}{l}{\textit{Panel A: One isolated task, diving}}                & &    \\
			\rule{0pt}{0.5ex} \\
              & Task 1:          & & Multiplier:       & & Combined:   \\
			& \multicolumn{1}{c}{Performance} & ~~~ & \multicolumn{1}{c}{Difficulty} & ~~~ & \multicolumn{1}{c}{Total score}\\
			\cline{2-2} \cline{4-4} \cline{6-6} \rule{0pt}{3ex} \\

			Positive Feedback   & 0.115***  & & -0.002      & & 1.071***  \\
			                    & (0.032)   & & (0.006)     & & (0.324)   \\
			Negative Feedback   &  0.001    & & -0.001      & & -0.029    \\
			                    & (0.029)   & & (0.005)     & & (0.282)   \\
			\rule{0pt}{3ex} \\
			\multicolumn{4}{l}{\textit{Panel B: Two simultaneous tasks, ski jumping}}                & &    \\
			\rule{0pt}{0.5ex} \\
            \multicolumn{1}{l}{} ~~~~~~~~~~~~~~~~~~~~~~~~~~~~~~~~~ & Task 1:          & & Task 2:       & & Combined:   \\

			& \multicolumn{1}{c}{Performance}   & ~~~ & \multicolumn{1}{c}{Distance points} & ~~~ & \multicolumn{1}{c}{Total score}\\
			
			\cline{2-2} \cline{4-4} \cline{6-6} \rule{0pt}{3ex} \\
			Positive Feedback   &  0.145*** & & 1.692***    & & 2.126***    \\
			                    & (0.034)   & & (0.634)     & & (0.693)     \\
			Negative Feedback   & -0.049    & &   0.072     & &   -0.080    \\
			                    & (0.037)   & &   (0.545)   & &  (0.631)    \\
			                    \rule{0pt}{0.1ex} \\
			\bottomrule
			\multicolumn{6}{l}{\footnotesize Notes: Linear Regression estimates. Control variables as in column (3) in Table \ref{tab:table_base}. Feedback was }\\
			\multicolumn{6}{l}{\footnotesize ~~~~~~~~~ given previously for Task 1 only. Standard errors are clustered on the individual level. *, ** }\\
			\multicolumn{6}{l}{\footnotesize ~~~~~~~~~ , and *** represents statistical significance at the 10 \%, 5 \%, and 1 \% level, respectively.}
		\end{tabular}
		
	\end{table}
	
	\subsection{Robustness}\label{sec:robu}
 
	To ensure that our results are robust to different specifications we conduct several supplementary analyses. First, we consider alternative specifications of our key variables. In a first regression, we take the mean of all (five or seven) judges' ratings, instead of the performance, i.e., the mean of the (after discarding the extreme ratings) remaining three ratings, as an alternative outcome variable. With the treatment, the second key variable is (additionally) constructed in two different ways. Instead of subtracting the jury's performance assessment from the most extreme (positive/negative) discarded rating we deduct (a) the lowest (highest) rating included in the jury's performance assessment from the lowest (highest) discarded rating (Deviation positive/negative$^+$) and (b) the jury's performance assessment from the mean of the two discarded highest or lowest ratings (Deviation positive/negative$^{++}$, in diving only). Table \ref{tab:table_robu} presents the results for these alternative specifications and shows robust estimates. We conclude from this that the result does neither depend on the concrete choice of the treatment variable, nor on the selection of the outcome variable.\par 
	
	Second, we consider different choices with respect to the sample that is used for the investigation. Data cleaning might offer some leeway to researchers influencing results. Thus, we provide additional analyses in Table \ref{tab:table_robu} using (a) the full sample without any data cleaning and (b) without excluding failed attempts (but excluding boundary values as described in Section \ref{sec:data_desc}). We find robust results for both supplementary analyses, indicating that our data-cleaning step does not drive the results.\par 
	
	Third, to prove that nationality bias is not responsible for the effect, i.e., judges favor their compatriots and potentially influence other judges on the panel, we re-estimate the results excluding all athletes with a compatriot judge in the panel. If the effect would be driven by these individuals the results might just be some mechanical effect. Though, the effect is also found for individuals not sharing nationality with a judge.\par

	\section{Managerial implications and conclusions}
	
    Giving feedback is one of the most important tasks of managers. On a typical workday, managers regularly provide feedback to their teams. Some of this feedback is subconscious, such as facial expressions or nodding as a sign of appreciation and approval. Other feedback can be formal and dictated by the institution, as is the case with appraisal interviews. It can be constructive and substantive. But it can also be purely motivational. Common examples would be phrases like “Good job!” or “You can do better!” embedded in the context of everyday conversations.\par

    The crucial question is whether such motivational feedback, given consciously by managers, can serve the goal of increasing the future productivity of workers. For both valences of feedback, i.e. positive and negative feedback, this question is not trivial. The appreciation that positive feedback expresses can motivate but also cause employees to rest on their laurels. Negative feedback can spur on but it can also hurt and discourage.\par

    Our causal analysis indicates that managers can use positive feedback to enhance productivity. Our results show a favorable impact of positive feedback on (subsequent) performance. The heterogeneity analysis indicates that this favorable effect of positive feedback can be found for feedback recipients coming from varying cultural backgrounds, for recipients of both male and female gender, and for relatively more and less experienced recipients. We find that the favorable effect of positive feedback is short-term, repeatable, and with potentially favorable spillover to related tasks. The favorable impact of positive feedback is robust to the setup in which the activity is performed and is more pronounced in highly relevant situations. All this makes us confident that giving positive motivational feedback is a performance-enhancing strategy.\par

    Furthermore, we find no significant impact of negative feedback on performance. This null effect might explain why managers and other raters are often reluctant to give negative feedback \cite{fisher1979transmission}, a phenomenon termed as leniency bias \cite{Cheng.2017} or MUM-effect \cite{Rosen.1970}. While in other contexts the lack of negative ratings is decreasing efficiency \cite{Cannon.2005, Bolton.2019, Keser.2021}, we report no need to give negative motivational feedback.\par

    Despite the robustness of our results, we acknowledge some limitations of our approach. First, our sample consists of internationally competing athletes. While their level of professionalism and self-discipline might be comparable to those of employees in highly competitive work environments, top athletes are not representative of the general population. Second, we consider an environment in which individuals receive feedback from multiple, external sources. Again, this is more comparable to daily life at large and competitive companies than at small firms. Third, we analyze a domain in which feedback recipients directly benefit from improvements in their performance, while feedback providers do not. In other domains, raters might be more prone to willfully bias their feedback. \par
    
    Therefore, we suggest that future research could contrast our results to environments, in which feedback providers benefit from an increased performance more than feedback recipients do. Employees in such environments might be prone to exploitation when employers use positive feedback as a substitute for more substantial improvements in the employees' well-being. Furthermore, future research could analyze the long-term effects of positive and negative feedback.\par
    
    With this study, we contribute to the literature that provides guidelines for optimal feedback \cite{Balcazar.1985, Alvero.2001, Sleiman.2020}. Our causal analysis shows that positive feedback is improving performance, while negative feedback has no effect.

@incollection{Villeval.2020,
 author = {Villeval, Marie Claire},
 title = {Performance Feedback and Peer Effects},
 pages = {1--38},
 publisher = {{Springer International Publishing}},
 isbn = {978-3-319-57365-6  },
 editor = {Zimmermann, Klaus F.},
 booktitle = {Handbook of Labor, Human Resources and Population Economics},
 year = {2020},
 address = {Cham},
 doi = {10.1007/978-3-319-57365-6_126-1  }
}

@article{Keser.2021,
 author = {Keser, Claudia and Sp{\"a}th, Maximilian},
 year = {2021},
 title = {The value of bad ratings: An experiment on the impact of distortions in reputation systems},
 pages = {101782},
 volume = {95},
 issn = {22148043},
 journal = {Journal of Behavioral and Experimental Economics},
 doi = {10.1016/j. socec.2021.101782 }
}

@article{Bolton.2019,
 author = {Bolton, Gary E. and Kusterer, David J. and Mans, Johannes},
 year = {2019},
 title = {Inflated Reputations: Uncertainty, Leniency, and Moral Wiggle Room in Trader Feedback Systems},
 pages = {5371--5391},
 volume = {65},
 number = {11},
 issn = {0025-1909},
 journal = {Management Science},
 doi = {10.1287/mnsc.2018.3191 }
}

@article{Choi.2018,
 author = {Choi, Eunju and Johnson, Douglas A. and Moon, Kwangsu and Oah, Shezeen},
 year = {2018},
 title = {Effects of Positive and Negative Feedback Sequence on Work Performance and Emotional Responses},
 pages = {97--115},
 volume = {38},
 number = {2-3},
 issn = {0160-8061},
 journal = {Journal of Organizational Behavior Management},
 doi = {10.1080/01608061.2017.1423151}
}

@article{Burgers.2015,
 author = {Burgers, Christian and Eden, Allison and {van Engelenburg}, M{\'e}lisande D. and Buningh, Sander},
 year = {2015},
 title = {How feedback boosts motivation and play in a brain-training game},
 pages = {94--103},
 volume = {48},
 issn = {07475632},
 journal = {Computers in Human Behavior},
 doi = {10.1016/j.chb.2015.01.038}
}

@article{Johnson.2015,
 author = {Johnson, Douglas A. and Rocheleau, Jessica M. and Tilka, Rachael E.},
 year = {2015},
 title = {Considerations in Feedback Delivery: The Role of Accuracy and Type of Evaluation},
 pages = {240--258},
 volume = {35},
 number = {3-4},
 issn = {0160-8061},
 journal = {Journal of Organizational Behavior Management},
 doi = {10.1080/01608061.2015.1093055}
}

@article{Alvero.2001,
 author = {Alvero, Alicia M. and Bucklin, Barbara R. and Austin, John},
 year = {2001},
 title = {An Objective Review of the Effectiveness and Essential Characteristics of Performance Feedback in Organizational Settings (1985-1998)},
 pages = {3--29},
 volume = {21},
 number = {1},
 issn = {0160-8061},
 journal = {Journal of Organizational Behavior Management},
 doi = {10.1300/J075v21n01_02}
}

@article{Balcazar.1985,
 author = {Balcazar, Fabricio and Hopkins, Bill L. and Suarez, Yolanda},
 year = {1985},
 title = {A Critical, Objective Review of Performance Feedback},
 pages = {65--89},
 volume = {7},
 number = {3-4},
 issn = {0160-8061},
 journal = {Journal of Organizational Behavior Management},
 doi = {10.1300/J075v07n03_05 }
}

@article{Sleiman.2020,
 author = {Sleiman, Andressa A. and Sigurjonsdottir, Sigridur and Elnes, Aud and Gage, Nicholas A. and Gravina, Nicole E.},
 year = {2020},
 title = {A Quantitative Review of Performance Feedback in Organizational Settings (1998-2018)},
 pages = {303--332},
 volume = {40},
 number = {3-4},
 issn = {0160-8061},
 journal = {Journal of Organizational Behavior Management},
 doi = {10.1080/01608061.2020.1823300 }
}

@article{Swift.2018,
 abstract = {Although performance feedback is widely employed as a means to improve motivation, the efficacy and reliability of performance feedback is often obscured by individual differences and situational variables. The joint role of these moderating variables remains unknown. Accordingly, we investigate how the motivational impact of feedback is moderated by personality and task-difficulty. Utilizing three samples (total N = 916), we explore how Big Five personality traits moderate the motivational impact of false positive and negative feedback on playful, neutral, and frustrating puzzle tasks, respectively. Conscientious and Neurotic individuals together appear particularly sensitive to task difficulty, becoming significantly more motivated by negative feedback on playful tasks and demotivated by negative feedback on frustrating tasks. Results are discussed in terms of Goal-Setting and Self Determination Theory. Implications for industry and education are considered.},
 author = {Swift, Victor and Peterson, Jordan B.},
 year = {2018},
 title = {Improving the effectiveness of performance feedback by considering personality traits and task demands},
 pages = {e0197810},
 volume = {13},
 number = {5},
 journal = {PloS one},
 doi = {10.1371/journal.pone.0197810 },
 file = {http://www.ncbi.nlm.nih.gov/pubmed/29787593},
 file = {https://www.ncbi.nlm.nih.gov/pmc/articles/PMC5963754}
}

@article{Waldersee.1994,
 author = {Waldersee, Robert and Luthans, Fred},
 year = {1994},
 title = {The impact of positive and corrective feedback on customer service performance},
 pages = {83--95},
 volume = {15},
 number = {1},
 issn = {08943796},
 journal = {Journal of Organizational Behavior},
 doi = {10.1002/job.4030150109 }
}

@article{Vancouver.2004,
 abstract = {Control theories claim that information about performance is often used by multiple goal systems. A proposition tested here was that performance information can create discrepancies in self-concept goals, directing cognitive resources away from the task goal system. To manipulate performance information, 160 undergraduates were given false positive or false negative normative feedback while working on a task that did or did not require substantial cognitive resources. Half of the participants were then given an opportunity to reaffirm their self-concepts following feedback, whereas half were not. Feedback sign positively related to performance only for those working on the cognitively intense task and not given a chance to reaffirm. Otherwise, feedback sign was negatively related to performance, albeit weakly.},
 author = {Vancouver, Jeffrey B. and Tischner, E. Casey},
 year = {2004},
 title = {The effect of feedback sign on task performance depends on self-concept discrepancies},
 pages = {1092--1098},
 volume = {89},
 number = {6},
 issn = {0021-9010},
 journal = {The Journal of Applied Psychology},
 doi = {10.1037/0021-9010.89.6.1092  },
 file = {http://www.ncbi.nlm.nih.gov/pubmed/15584844}
}

@article{Podsakoff.1989,
 author = {Podsakoff, Philip M. and Farh, Jiing-Lih},
 year = {1989},
 title = {Effects of feedback sign and credibility on goal setting and task performance},
 pages = {45--67},
 volume = {44},
 number = {1},
 issn = {07495978},
 journal = {Organizational Behavior and Human Decision Processes},
 doi = {10.1016/0749-5978(89)90034-4  }
}

@article{Lechermeier.2018,
 author = {Lechermeier, Jonas and Fassnacht, Martin},
 year = {2018},
 title = {How do performance feedback characteristics influence recipients' reactions? A state-of-the-art review on feedback source, timing, and valence effects},
 pages = {145--193},
 volume = {68},
 number = {2},
 issn = {2198-1620},
 journal = {Management Review Quarterly},
 doi = {10.1007/s11301-018-0136-8 }
}

@article{Dellarocas.2008,
 author = {Dellarocas, Chrysanthos and Wood, Charles A.},
 year = {2008},
 title = {The Sound of Silence in Online Feedback: Estimating Trading Risks in the Presence of Reporting Bias},
 pages = {460--476},
 volume = {54},
 number = {3},
 issn = {0025-1909},
 journal = {Management Science},
 doi = {10.1287/mnsc.1070.0747}
}

@article{Azmat.2019,
 author = {Azmat, Ghazala and Bagues, Manuel and Cabrales, Antonio and Iriberri, Nagore},
 year = {2019},
 title = {What You Don't Know$\ldots$Can't Hurt You? A Natural Field Experiment on Relative Performance Feedback in Higher Education},
 pages = {3714--3736},
 volume = {65},
 number = {8},
 issn = {0025-1909},
 journal = {Management Science},
 doi = {10.1287/mnsc.2018.3131}
}

@article{Kuhnen.2012,
 author = {Kuhnen, Camelia M. and Tymula, Agnieszka},
 year = {2012},
 title = {Feedback, Self-Esteem, and Performance in Organizations},
 pages = {94--113},
 volume = {58},
 number = {1},
 issn = {0025-1909},
 journal = {Management Science},
 doi = {10.1287/mnsc.1110.1379}
}

% This file was created with Citavi 6.8.0.0

@article{Swift.2018,
 abstract = {Although performance feedback is widely employed as a means to improve motivation, the efficacy and reliability of performance feedback is often obscured by individual differences and situational variables. The joint role of these moderating variables remains unknown. Accordingly, we investigate how the motivational impact of feedback is moderated by personality and task-difficulty. Utilizing three samples (total N = 916), we explore how Big Five personality traits moderate the motivational impact of false positive and negative feedback on playful, neutral, and frustrating puzzle tasks, respectively. Conscientious and Neurotic individuals together appear particularly sensitive to task difficulty, becoming significantly more motivated by negative feedback on playful tasks and demotivated by negative feedback on frustrating tasks. Results are discussed in terms of Goal-Setting and Self Determination Theory. Implications for industry and education are considered.},
 author = {Swift, Victor and Peterson, Jordan B.},
 year = {2018},
 title = {Improving the effectiveness of performance feedback by considering personality traits and task demands},
 pages = {e0197810},
 volume = {13},
 number = {5},
 journal = {PloS one},
 doi = {10.1371/journal.pone.0197810},
 file = {http://www.ncbi.nlm.nih.gov/pubmed/29787593},
 file = {https://www.ncbi.nlm.nih.gov/pmc/articles/PMC5963754}
}

@article{Lechermeier.2018,
 author = {Lechermeier, Jonas and Fassnacht, Martin},
 year = {2018},
 title = {How do performance feedback characteristics influence recipients' reactions? A state-of-the-art review on feedback source, timing, and valence effects},
 pages = {145--193},
 volume = {68},
 number = {2},
 issn = {2198-1620},
 journal = {Management Review Quarterly},
 doi = {10.1007/s11301-018-0136-8}
}

@article{Kuhnen.2012,
 author = {Kuhnen, Camelia M. and Tymula, Agnieszka},
 year = {2012},
 title = {Feedback, Self-Esteem, and Performance in Organizations},
 pages = {94--113},
 volume = {58},
 number = {1},
 issn = {0025-1909},
 journal = {Management Science},
 doi = {10.1287/mnsc.1110.1379}
}

@article{Johnson.2015,
 author = {Johnson, Douglas A. and Rocheleau, Jessica M. and Tilka, Rachael E.},
 year = {2015},
 title = {Considerations in Feedback Delivery: The Role of Accuracy and Type of Evaluation},
 pages = {240--258},
 volume = {35},
 number = {3-4},
 issn = {0160-8061},
 journal = {Journal of Organizational Behavior Management},
 doi = {10.1080/01608061.2015.1093055}
}

@article{Eil.2011,
 author = {Eil, David and Rao, Justin M.},
 year = {2011},
 title = {The Good News-Bad News Effect: Asymmetric Processing of Objective Information about Yourself},
 pages = {114--138},
 volume = {3},
 number = {2},
 issn = {1945-7669},
 journal = {American Economic Journal: Microeconomics},
 doi = {10.1257/mic.3.2.114}
}

@article{Dellarocas.2008,
 author = {Dellarocas, Chrysanthos and Wood, Charles A.},
 year = {2008},
 title = {The Sound of Silence in Online Feedback: Estimating Trading Risks in the Presence of Reporting Bias},
 pages = {460--476},
 volume = {54},
 number = {3},
 issn = {0025-1909},
 journal = {Management Science},
 doi = {10.1287/mnsc.1070.0747}
}

@article{Chadd.2021,
 author = {Chadd, Ian and Filiz-Ozbay, Emel and Ozbay, Erkut Y.},
 year = {2021},
 title = {The relevance of irrelevant information},
 pages = {985--1018},
 volume = {24},
 number = {3},
 issn = {1386-4157},
 journal = {Experimental Economics},
 doi = {10.1007/s10683-020-09687-3}
}

@article{Azmat.2019,
 author = {Azmat, Ghazala and Bagues, Manuel and Cabrales, Antonio and Iriberri, Nagore},
 year = {2019},
 title = {What You Don't Know$\ldots$Can't Hurt You? A Natural Field Experiment on Relative Performance Feedback in Higher Education},
 pages = {3714--3736},
 volume = {65},
 number = {8},
 issn = {0025-1909},
 journal = {Management Science},
 doi = {10.1287/mnsc.2018.3131}
}

@article{Chadd.2021,
 author = {Chadd, Ian and Filiz-Ozbay, Emel and Ozbay, Erkut Y.},
 year = {2021},
 title = {The relevance of irrelevant information},
 pages = {985--1018},
 volume = {24},
 number = {3},
 issn = {1386-4157},
 journal = {Experimental Economics},
 doi = {10.1007/s10683-020-09687-3}
}

@article{ca1c829d041c4dcb8d8880e9989329b8,
title = "Partner selection supported by opaque reputation promotes cooperative behavior",
abstract = "Reputation plays a major role in human societies, and it has been proposed as an explanation for the evolution of cooperation.While the majority of previous studies equates reputation with a transparent and complete history of players{\textquoteright} past decisions, reputations in real life are often ambiguous and opaque. Using web-based experiments, we explore the extent to which opaque reputation works in isolating defectors, with and without partner selection opportunities. We found that low reputation worksas a signal of untrustworthiness, whereas medium or high reputations are not taken into account by subjects for orienting their choices. Reputation without partner selection does not promote cooperative behavior; that is, defectors do not turn into cooperators only for the sake of getting a positive reputation. Finally, in a third study, when reputation is pivotal to selection,then a substantial proportion of would-be-defectors turn into cooperators. Taken together, these results provide insights about the characteristics of reputation and about the way in which humans make use of it when selecting partners, and also when knowing that they will be selected.",
keywords = "reputation, Partner selection, cooperation, prisoner's dilemma, online transactions",
author = "Valerio Capraro and Francesca Giardini and Daniele Vilone and Mario Paolucci",
year = "2016",
month = nov,
language = "English",
volume = "11",
pages = "589--600",
journal = "Judgement and Decision Making",
issn = "1930-2975",
publisher = "SOC JUDGMENT & DECISION MAKING",
number = "6",

}

@article{Baumeister.2001,
 author = {Baumeister, Roy F. and Bratslavsky, Ellen and Finkenauer, Catrin and Vohs, Kathleen D.},
 year = {2001},
 title = {Bad is Stronger than Good},
 pages = {323--370},
 volume = {5},
 number = {4},
 issn = {1089-2680},
 journal = {Review of General Psychology},
 doi = {10.1037/1089-2680.5.4.323}
}

@article{Luckingreiley.2007,
 author = {Lucking-Reiley, David and Bryan, Doug and Prasad, Naghi and Reeves, Daniel},
 year = {2007},
 title = {PENNIES FROM EBAY: THE DETERMINANTS OF PRICE IN ONLINE AUCTIONS},
 pages = {223--233},
 volume = {55},
 number = {2},
 issn = {0022-1821},
 journal = {Journal of Industrial Economics},
 doi = {10.1111/j.1467-6451.2007.00309.x}
}

@article{Standifird.2001,
 author = {Standifird, S.},
 year = {2001},
 title = {Reputation and e-commerce: eBay auctions and the asymmetrical impact of positive and negative ratings},
 pages = {279--295},
 volume = {27},
 number = {3},
 issn = {01492063},
 journal = {Journal of Management},
 doi = {10.1016/S0149-2063(01)00092-7 }
}

@article{Watson.1969,
 author = {Watson, D. and Friend, R.},
 year = {1969},
 title = {Measurement of social-evaluative anxiety},
 pages = {448--457},
 volume = {33},
 number = {4},
 issn = {0022-006X},
 journal = {Journal of consulting and clinical psychology},
 doi = {10.1037/h0027806  },
 file = {http://www.ncbi.nlm.nih.gov/pubmed/5810590}
}

@article{Morgeson.2010,
 author = {Morgeson, Frederick P. and DeRue, D. Scott and Karam, Elizabeth P.},
 year = {2010},
 title = {Leadership in Teams: A Functional Approach to Understanding Leadership Structures and Processes},
 pages = {5--39},
 volume = {36},
 number = {1},
 issn = {01492063},
 journal = {Journal of Management},
 doi = {10.1177/0149206309347376  }
}

@article{Krumer.2022,
  title={Nationalistic bias among international experts: Evidence from professional ski jumping},
  author={Krumer, Alex and Otto, Felix and Pawlowski, Tim},
  journal={The Scandinavian Journal of Economics},
  volume={124},
  number={1},
  pages={278--300},
  year={2022},
  publisher={Wiley Online Library},
  doi={10.1111/sjoe.12451}
}

@article{Heiniger.2021,
  title={Judging the judges: evaluating the accuracy and national bias of international gymnastics judges},
  author={Heiniger, Sandro and Mercier, Hugues},
  journal={Journal of Quantitative Analysis in Sports},
  volume={17},
  number={4},
  pages={289--305},
  year={2021},
  publisher={De Gruyter},
  doi={10.1515/jqas-2019-0113 }
}

@article{Kahn.2000,
  title={The sports business as a labor market laboratory},
  author={Kahn, Lawrence M},
  journal={Journal of economic perspectives},
  volume={14},
  number={3},
  pages={75--94},
  year={2000}
}

@article{Levitt.2008,
  title={Homo economicus evolves},
  author={Levitt, Steven D and List, John A},
  journal={Science},
  volume={319},
  number={5865},
  pages={909--910},
  year={2008},
  publisher={American Association for the Advancement of Science}
}

@article{Bartling.2015,
  title={Expectations as reference points: Field evidence from professional soccer},
  author={Bartling, Bj{\"o}rn and Brandes, Leif and Schunk, Daniel},
  journal={Management Science},
  volume={61},
  number={11},
  pages={2646--2661},
  year={2015},
  publisher={INFORMS}
}

@article{Massey.2013,
  title={The loser's curse: Decision making and market efficiency in the National Football League draft},
  author={Massey, Cade and Thaler, Richard H},
  journal={Management Science},
  volume={59},
  number={7},
  pages={1479--1495},
  year={2013},
  publisher={INFORMS}
}

@article{Pope.2018,
  title={Awareness reduces racial bias},
  author={Pope, Devin G and Price, Joseph and Wolfers, Justin},
  journal={Management Science},
  volume={64},
  number={11},
  pages={4988--4995},
  year={2018},
  publisher={INFORMS}
}

@article{Henrich.2010,
 author = {Henrich, Joseph and Heine, Steven J. and Norenzayan, Ara},
 year = {2010},
 title = {Most people are not WEIRD},
 pages = {29},
 volume = {466},
 number = {7302},
 journal = {Nature},
 doi = {10.1038/466029a },
 file = {http://www.ncbi.nlm.nih.gov/pubmed/20595995}
}

@article{Bond.1987,
 author = {Bond, Charles F. and Anderson, Evan L.},
 year = {1987},
 title = {The reluctance to transmit bad news: Private discomfort or public display?},
 pages = {176--187},
 volume = {23},
 number = {2},
 issn = {00221031},
 journal = {Journal of Experimental Social Psychology},
 doi = {10.1016/0022-1031(87)90030-8 }
}

@article{Cheng.2017,
 abstract = {Some researchers assume that employees' personality characteristics affect leniency in rating others and themselves. However, little research has investigated these two tendencies at the same time. In the present study we developed one index for other-rating leniency and another one for self-rating leniency. Based on a review of the literature, we hypothesized that a generous assessment of peers would more likely be made by those who are extroverted and agreeable than by those who are not. Furthermore, a generous assessment of oneself would more likely be made by people who are conscientious and emotionally stable, than by people who are not. We also investigated if the leniency in rating others and the leniency in rating oneself are part of a more general leniency tendency. Data collected from a sample of real estate dealers provided support for the above hypotheses. Limitations and implications for future research are discussed.},
 author = {Cheng, Kevin H. C. and Hui, C. Harry and Cascio, Wayne F.},
 year = {2017},
 title = {Leniency Bias in Performance Ratings: The Big-Five Correlates},
 pages = {521},
 volume = {8},
 issn = {1664-1078},
 journal = {Frontiers in Psychology},
 doi = {10.3389/fpsyg.2017.00521},
 file = {http://www.ncbi.nlm.nih.gov/pubmed/28443043},
 file = {https://www.ncbi.nlm.nih.gov/pmc/articles/PMC5385382}
}

@article{Larson.1986,
 author = {Larson, James R.},
 year = {1986},
 title = {Supervisors' performance feedback to subordinates: The impact of subordinate performance valence and outcome dependence},
 pages = {391--408},
 volume = {37},
 number = {3},
 issn = {07495978},
 journal = {Organizational Behavior and Human Decision Processes},
 doi = {10.1016/0749-5978(86)90037-3 }
}

@article{Rosen.1970,
 author = {Rosen, Sidney and Tesser, Abraham},
 year = {1970},
 title = {On Reluctance to Communicate Undesirable Information: The MUM Effect},
 pages = {253},
 volume = {33},
 number = {3},
 issn = {00380431},
 journal = {Sociometry},
 doi = {10.2307/2786156  }
}

@article{Weidinger.2016,
 author = {Weidinger, Anne F. and Spinath, Birgit and Steinmayr, Ricarda},
 year = {2016},
 title = {Why does intrinsic motivation decline following negative feedback? The mediating role of ability self-concept and its moderation by goal orientations},
 pages = {117--128},
 volume = {47},
 issn = {10416080},
 journal = {Learning and Individual Differences},
 doi = {10.1016/j.  lindif.2016.01.003}
}

@article{Fong.2019,
 author = {Fong, Carlton J. and Patall, Erika A. and Vasquez, Ariana C. and Stautberg, Sandra},
 year = {2019},
 title = {A Meta-Analysis of Negative Feedback on Intrinsic Motivation},
 pages = {121--162},
 volume = {31},
 number = {1},
 issn = {1040-726X},
 journal = {Educational Psychology Review},
 doi = {10.1007/s10648-018-9446-6 }
}

@article{Markus.1991,
  title={Culture and the self: Implications for cognition, emotion, and motivation.},
  author={Markus, Hazel R and Kitayama, Shinobu},
  journal={Psychological Review},
  volume={98},
  number={2},
  pages={224},
  year={1991},
  publisher={American Psychological Association},
  doi={10.1037/0033-295X.98.2.224}
}

@article{Vallerand.1988,
 author = {Vallerand, Robert J. and Reid, Greg},
 year = {1988},
 title = {On the relative effects of positive and negative verbal feedback on males' and females' intrinsic motivation},
 pages = {239--250},
 volume = {20},
 number = {3},
 issn = {0008-400X},
 journal = {Canadian Journal of Behavioural Science / Revue canadienne des sciences du comportement},
 doi = {10.1037/h0079930}
}

@misc{Deci.1972,
 author = {Deci, Eward L. and Casico, Wayne F.},
 year = {1972},
 title = {Changes in Intrinsic Motivation as a Function of Negative Feedback and Threats.},
 journal = {Paper presented at the Eastern Psychological Association Meeting in Boston, MA.}
}

@article{Fishbach.2010,
 author = {Fishbach, Ayelet and Eyal, Tal and Finkelstein, Stacey R.},
 year = {2010},
 title = {How Positive and Negative Feedback Motivate Goal Pursuit},
 pages = {517--530},
 volume = {4},
 number = {8},
 issn = {17519004},
 journal = {Social and Personality Psychology Compass},
 doi = {10.1111/j.1751-9004.2010.00285.x}
}

@article{Itzchakov.2020,
 author = {Itzchakov, Guy and Latham, Gary P.},
 year = {2020},
 title = {The Moderating Effect of Performance Feedback and the Mediating Effect of Self--Set Goals on the Primed Goal--Performance Relationship},
 pages = {379--414},
 volume = {69},
 number = {2},
 issn = {0269-994X},
 journal = {Applied Psychology},
 doi = {10.1111/apps.12176  }
}

@article{Carpentier.2013,
 author = {Carpentier, Jo{\"e}lle and Mageau, Genevi{\`e}ve A.},
 year = {2013},
 title = {When change-oriented feedback enhances motivation, well-being and performance: A look at autonomy-supportive feedback in sport},
 pages = {423--435},
 volume = {14},
 number = {3},
 issn = {14690292},
 journal = {Psychology of Sport and Exercise},
 doi = {10.1016/j.  psychsport.2013.01.003}
}

@article{Emerson.2009,
  title={Assessing judging bias: An example from the 2000 Olympic Games},
  author={Emerson, John W and Seltzer, Miki and Lin, David},
  journal={The American Statistician},
  volume={63},
  number={2},
  pages={124--131},
  year={2009},
  publisher={Taylor \& Francis}
}

@article{Sandberg.2018,
  title={Competing identities: a field study of in-group bias among professional evaluators},
  author={Sandberg, Anna},
  journal={The Economic Journal},
  volume={128},
  number={613},
  pages={2131--2159},
  year={2018},
  publisher={Wiley Online Library},
  doi={10.1111/ecoj.12513 }
}

@article{Morgan.2014,
  title={The harder the task, the higher the score: Findings of a difficulty bias},
  author={Morgan, Hillary N and Rotthoff, Kurt W},
  journal={Economic Inquiry},
  volume={52},
  number={3},
  pages={1014--1026},
  year={2014},
  publisher={Wiley Online Library},
  doi={10.1111/ecin.12074 }
}

@article{Damisch.2006,
  title={Olympic medals as fruits of comparison? Assimilation and contrast in sequential performance judgments.},
  author={Damisch, Lysann and Mussweiler, Thomas and Plessner, Henning},
  journal={Journal of Experimental Psychology: Applied},
  volume={12},
  number={3},
  pages={166},
  year={2006},
  publisher={American Psychological Association}
}

@book{Bar.2011,
  title={Judgment, decision-making and success in sport},
  author={Bar-Eli, Michael and Plessner, Henning and Raab, Markus},
  volume={1},
  year={2011},
  publisher={John Wiley \& Sons}
}

@article{Findlay.2004,
  title={A reputation bias in figure skating judging},
  author={Findlay, Leanne C and Ste-Marie, Diane M},
  journal={Journal of Sport and Exercise Psychology},
  volume={26},
  number={1},
  pages={154--166},
  year={2004},
  publisher={Citeseer},
  doi={10.1123/jsep.26.1.154 }
}

@article{Zitzewitz.2006,
  title={Nationalism in winter sports judging and its lessons for organizational decision making},
  author={Zitzewitz, Eric},
  journal={Journal of Economics \& Management Strategy},
  volume={15},
  number={1},
  pages={67--99},
  year={2006},
  publisher={Wiley Online Library}
}

@article{Ginsburgh.2003,
  title={Expert opinion and compensation: Evidence from a musical competition},
  author={Ginsburgh, Victor A and Van Ours, Jan C},
  journal={The American Economic Review},
  volume={93},
  number={1},
  pages={289--296},
  year={2003},
  publisher={JSTOR},
  doi={10.1257/000282803321455296 }
}

@article{vanDijk.2011,
 author = {{van Dijk}, Dina and Kluger, Avraham N.},
 year = {2011},
 title = {Task type as a moderator of positive/negative feedback effects on motivation and performance: A regulatory focus perspective},
 pages = {1084--1105},
 volume = {32},
 number = {8},
 issn = {08943796},
 journal = {Journal of Organizational Behavior},
 doi = {10.1002/job.725 }
}

@article{Kluger.1996,
 author = {Kluger, Avraham N. and DeNisi, Angelo},
 year = {1996},
 title = {The effects of feedback interventions on performance: A historical review, a meta-analysis, and a preliminary feedback intervention theory},
 pages = {254--284},
 volume = {119},
 number = {2},
 issn = {0033-2909},
 journal = {Psychological Bulletin},
 doi = {10.1037/0033-2909.119.2.254  }
}

% This file was created with Citavi 6.8.0.0

@article{Su.2022,
 abstract = {As an important tool for supervisors to intervene subordinates' work and influence their performance, supervisor feedback has gradually become a new academic research hotspot. In this study, we build and verify a theoretical model to explore the different effects of supervisor positive and negative feedback on subordinate in-role and extra-role performance, and the moderating role of regulatory focus in these relationships. With data from pairing samples of 403 Chinese employees and their direct supervisors, the results indicate that supervisor positive feedback is positively related to subordinate in-role and extra-role performance. Supervisor negative feedback is positively related to subordinate in-role performance and negatively related to subordinate extra-role performance. Regulatory focus of subordinate can moderate the influence of supervisor positive feedback on subordinate in-role and extra-role performance, but it cannot moderate the influence of supervisor negative feedback on subordinate in-role and extra-role performance. That means when subordinates have promotion focus, the influence of supervisor positive feedback on their in-role performance and extra-role performance was stronger than those with prevention focus. These results further enrich the research on the relationship between supervisor feedback and subordinate performance, especially the different effects of positive and negative feedback from supervisor on subordinate with different regulatory focus. All conclusions from the analyses above not only further verify and develop some previous points on supervisor feedback and subordinate performance, but also derive certain management implications for promoting subordinate in-role and extra-role performance from the perspective of supervisor positive and negative feedback.},
 author = {Su, Weilin and Yuan, Shuai and Qi, Qian},
 year = {2022},
 title = {Different Effects of Supervisor Positive and Negative Feedback on Subordinate In-Role and Extra-Role Performance: The Moderating Role of Regulatory Focus},
 pages = {757687},
 volume = {12},
 issn = {1664-1078},
 journal = {Frontiers in Psychology},
 doi = {10.3389/fpsyg.2021.757687 },
 file = {http://www.ncbi.nlm.nih.gov/pubmed/35069334},
 file = {https://www.ncbi.nlm.nih.gov/pmc/articles/PMC8776992 }
}

@article{Taylor.1994,
 author = {Taylor, Jim and Demick, Andrew},
 year = {1994},
 title = {A multidimensional model of momentum in sports},
 pages = {51--70},
 volume = {6},
 number = {1},
 issn = {1041-3200},
 journal = {Journal of Applied Sport Psychology},
 doi = {10.1080/10413209408406465    }
}

@article{Zizzo.2010,
 author = {Zizzo, Daniel John},
 year = {2010},
 title = {Experimenter demand effects in economic experiments},
 pages = {75--98},
 volume = {13},
 number = {1},
 issn = {1386-4157},
 journal = {Experimental Economics},
 doi = {10.1007/s10683-009-9230-z    }
}

@article{Stone.1983,
  title={The Effects of Feedback Favorability and Feedback Consistency.},
  author={Stone, Dianna L and Stone, Eugene F},
  journal={Academy of Management Proceedings},
  volume={1983},
  number={1},
  pages={178--182},
  year={1983},
  doi={10.1016/0749-5978(85)90011-1 }
}

@article{Bear.2017,
  title={Performance feedback, power retention, and the gender gap in leadership},
  author={Bear, Julia B and Cushenbery, Lily and London, Manuel and Sherman, Gary D},
  journal={The Leadership Quarterly},
  volume={28},
  number={6},
  pages={721--740},
  year={2017},
  publisher={Elsevier}
}

@article{Moore.2008,
  title={The trouble with overconfidence.},
  author={Moore, Don A and Healy, Paul J},
  journal={Psychological Review},
  volume={115},
  number={2},
  pages={502},
  year={2008},
  publisher={American Psychological Association},
  doi={10.1037/0033-295X.115.2.502  }
}

@article{Goller.2022,
author = {Goller, Daniel},
doi = {10.1007/s10479-022-04563-0     },
journal = {Annals of Operations Research},
title = {{Analysing a built-in advantage in asymmetric darts contests using causal machine learning}},
volume = {forthcom.},
pages = {1--31},
year = {2022}
}
@article{Page.2010,
  title={Alone against the crowd: Individual differences in referees’ ability to cope under pressure},
  author={Page, Katie and Page, Lionel},
  journal={Journal of Economic Psychology},
  volume={31},
  number={2},
  pages={192--199},
  year={2010},
  publisher={Elsevier},
  doi={10.1016/j.joep.2009.08.007}
}

@article{Goller.2020,
  title={Let's meet as usual: Do games played on non-frequent days differ? Evidence from top European soccer leagues},
  author={Goller, Daniel and Krumer, Alex},
  journal={European Journal of Operational Research},
  volume={286},
  number={2},
  pages={740--754},
  year={2020},
  publisher={Elsevier},
  doi={10.1016/j.    ejor.2020.03.062}
}

@article{GH.2022,
  title={A general framework to quantify the event importance in multi-event contests},
  author={Goller, Daniel and Heiniger, Sandro},
  journal={arXiv preprint arXiv:2207.02316},
  year={2022}
}

@article{Pulford.1997,
  title={Overconfidence: Feedback and item difficulty effects},
  author={Pulford, Briony D and Colman, Andrew M},
  journal={Personality and Individual Differences},
  volume={23},
  number={1},
  pages={125--133},
  year={1997},
  publisher={Elsevier},
  doi={10.1016/S0191-8869(97)00028-7 }
}

@article{Johnson.2013,
 author = {Johnson, Douglas A.},
 year = {2013},
 title = {A Component Analysis of the Impact of Evaluative and Objective Feedback on Performance},
 pages = {89--103},
 volume = {33},
 number = {2},
 issn = {0160-8061},
 journal = {Journal of Organizational Behavior Management},
 doi = {10.1080/01608061.2013.785879   }
}

@article{Bitchener.2008,
 author = {Bitchener, John},
 year = {2008},
 title = {Evidence in support of written corrective feedback},
 pages = {102--118},
 volume = {17},
 number = {2},
 issn = {10603743},
 journal = {Journal of Second Language Writing},
 doi = {10.1016/j.  jslw.2007.11.004}
}

@article{Ertac.2011,
 author = {Ertac, Seda},
 year = {2011},
 title = {Does self-relevance affect information processing? Experimental evidence on the response to performance and non-performance feedback},
 pages = {532--545},
 volume = {80},
 number = {3},
 issn = {01672681},
 journal = {Journal of Economic Behavior {\&} Organization},
 doi = {10.1016/j. jebo.2011.05.012}
}

@article{Barron.2021,
 author = {Barron, Kai},
 year = {2021},
 title = {Belief updating: does the `good-news, bad-news' asymmetry extend to purely financial domains?},
 pages = {31--58},
 volume = {24},
 number = {1},
 issn = {1386-4157},
 journal = {Experimental Economics},
 doi = {10.1007/s10683-020-09653-z }
}

@article{Kennedy.2017,
  title={Non-parametric methods for doubly robust estimation of continuous treatment effects},
  author={Kennedy, Edward H and Ma, Zongming and McHugh, Matthew D and Small, Dylan S},
  journal={Journal of the Royal Statistical Society: Series B (Statistical Methodology)},
  volume={79},
  number={4},
  pages={1229--1245},
  year={2017},
  publisher={Wiley Online Library},
  doi={10.1111/rssb.12212}
}

@article{Coutts.2019,
 author = {Coutts, Alexander},
 year = {2019},
 title = {Good news and bad news are still news: experimental evidence on belief updating},
 pages = {369--395},
 volume = {22},
 number = {2},
 issn = {1386-4157},
 journal = {Experimental Economics},
 doi = {10.1007/s10683-018-9572-5  }
}

@article{Kuzmanovic.2015,
 author = {Kuzmanovic, Bojana and Jefferson, Anneli and Vogeley, Kai},
 year = {2015},
 title = {Self-specific Optimism Bias in Belief Updating Is Associated with High Trait Optimism},
 pages = {281--293},
 volume = {28},
 number = {3},
 issn = {08943257},
 journal = {Journal of Behavioral Decision Making},
 doi = {10.1002/bdm.1849}
}

@article{Mobius.2022,
 author = {M{\"o}bius, Markus M. and Niederle, Muriel and Niehaus, Paul and Rosenblat, Tanya S.},
 year = {2022},
 title = {Managing Self-Confidence: Theory and Experimental Evidence},
 issn = {0025-1909},
 journal = {Management Science},
 doi = {10.1287/mnsc.2021.4294  }
}

@article{Sharot.2012,
 abstract = {Humans form beliefs asymmetrically; we tend to discount bad news but embrace good news. This reduced impact of unfavorable information on belief updating may have important societal implications, including the generation of financial market bubbles, ill preparedness in the face of natural disasters, and overly aggressive medical decisions. Here, we selectively improved people's tendency to incorporate bad news into their beliefs by disrupting the function of the left (but not right) inferior frontal gyrus using transcranial magnetic stimulation, thereby eliminating the engrained {\textquotedbl}good news/bad news effect.{\textquotedbl} Our results provide an instance of how selective disruption of regional human brain function paradoxically enhances the ability to incorporate unfavorable information into beliefs of vulnerability.},
 author = {Sharot, Tali and Kanai, Ryota and Marston, David and Korn, Christoph W. and Rees, Geraint and Dolan, Raymond J.},
 year = {2012},
 title = {Selectively altering belief formation in the human brain},
 pages = {17058--17062},
 volume = {109},
 number = {42},
 journal = {Proceedings of the National Academy of Sciences of the United States of America},
 doi = {10.1073/pnas.1205828109  },
 file = {http://www.ncbi.nlm.nih.gov/pubmed/23011798},
 file = {https://www.ncbi.nlm.nih.gov/pmc/articles/PMC3479523  }
}

@article{Azmat.2010,
 author = {Azmat, Ghazala and Iriberri, Nagore},
 year = {2010},
 title = {The importance of relative performance feedback information: Evidence from a natural experiment using high school students},
 pages = {435--452},
 volume = {94},
 number = {7-8},
 issn = {00472727},
 journal = {Journal of Public Economics},
 doi = {10.1016/j.jpubeco.2010.04.001 }
}

@article{Bandiera.2015,
 author = {Bandiera, Oriana and Larcinese, Valentino and Rasul, Imran},
 year = {2015},
 title = {Blissful ignorance? A natural experiment on the effect of feedback on students' performance},
 pages = {13--25},
 volume = {34},
 issn = {09275371},
 journal = {Labour Economics},
 doi = {10.1016/j.labeco.2015.02.002 }
}

@article{Abeler.2011,
 author = {Abeler, Johannes and Falk, Armin and Goette, Lorenz and Huffman, David},
 year = {2011},
 title = {Reference Points and Effort Provision},
 pages = {470--492},
 volume = {101},
 number = {2},
 issn = {0002-8282},
 journal = {American Economic Review},
 doi = {10.1257/aer.101.2.470 }
}

@article{Atwater.2007,
 author = {Atwater, Leanne E. and Brett, Joan F. and Charles, Atira Cherise},
 year = {2007},
 title = {Multisource feedback: Lessons learned and implications for practice},
 pages = {285--307},
 volume = {46},
 number = {2},
 issn = {00904848},
 journal = {Human Resource Management},
 doi = {10.1002/hrm.20161  }
}

@article{Bailey.2002,
 author = {Bailey, Caroline and Fletcher, Clive},
 year = {2002},
 title = {The impact of multiple source feedback on management development: findings from a longitudinal study},
 pages = {853--867},
 volume = {23},
 number = {7},
 issn = {08943796},
 journal = {Journal of Organizational Behavior},
 doi = {10.1002/job.167  }
}

@article{Cason.1998,
 abstract = {This paper introduces the sequential dictator game to study how social influence may affect subjects' choices when making dictator allocations. Subjects made dictator allocations of {\$}40 before and after learning the allocation made by one other subject in the Relevant Information treatment, or the birthday of one other subject in the Irrelevant Information treatment. Subjects on average become more self-regarding in the Irrelevant Information treatment, but observing relevant information constrains some subjects from moving toward more self-regarding choices. We also find that subjects who exhibit more self-regarding behavior on their first decisions are less likely to change choices between their first and second decisions, and the use of the Strategy Method in this setting does not significantly alter choices. The relationships between our findings and the economic and psychological literature regarding how social influence operates are also explored. Copyright 1998 Academic Press.},
 author = {Cason and Mui},
 year = {1998},
 title = {Social Influence in the Sequential Dictator Game},
 pages = {248--265},
 volume = {42},
 number = {2/3},
 issn = {0022-2496},
 journal = {Journal of Mathematical Psychology},
 doi = {10.1006/jmps.1998.1213 },
 file = {http://www.ncbi.nlm.nih.gov/pubmed/9710550}
}

@article{Stone.1984,
 author = {Stone, Eugene F. and Stone, Dianna L.},
 year = {1984},
 title = {The Effects of Multiple Sources of Performance Feedback and Feedback Favorability on Self-Perceived Task Competence and Perceived Feedback Accuracy},
 pages = {371--378},
 volume = {10},
 number = {3},
 issn = {01492063},
 journal = {Journal of Management},
 doi = {10.1177/014920638401000311  }
}

@article{Smither.2005,
 author = {Smither, James. W. and London, Manuel and Reilley, Richard R.},
 year = {2005},
 title = {Does Performance improve following Multisource Feedback? A theoretical Model, Meta-Analysis, and Review of Empirical Findings},
 pages = {33--66},
 volume = {58},
 number = {1},
 issn = {0031-5826},
 journal = {Personnel Psychology},
 doi = {10.1111/j.1744-6570.2005.514_1.x    }
}

@article{fisher1979transmission,
  title={Transmission of positive and negative feedback to subordinates: A laboratory investigation.},
  author={Fisher, Cynthia D},
  journal={Journal of Applied Psychology},
  volume={64},
  number={5},
  pages={533},
  year={1979},
  publisher={American Psychological Association},
  doi={10.1037/0021-9010.64.5.533  }
}

@article{Breiman.2001,
  title={Random forests},
  author={Breiman, Leo},
  journal={Machine Learning},
  volume={45},
  number={1},
  pages={5--32},
  year={2001},
  publisher={Springer},
  doi = {10.1023/A:1010933404324  }
}

@article{Budelmann.02.11.2022,
 author = {Budelmann, Jeannine},
 year = {02.11.2022},
 title = {{\frqq}Das haben Sie gut gemacht!{\flqq} - Wie Chefs die Kraft des Lobes ausnutzen},
 url = {https://www.spiegel.de/start/lob-bei-der-arbeit-warum-es-zuweilen-mit-vorsicht-zu-geniessen-ist-a-a1afea8d-59c8-4709-ae32-753fa2ec9a77?utm_source=pocket-newtab-global-de-DE},
 urldate = {21.11.2022},
 journal = {Spiegel Online}
}

%}

@article{HECHT.2012,
 author = {Hecht, Gary and Tafkov, Ivo and Towry, Kristy L.},
 year = {2012},
 title = {Performance Spillover in a Multitask Environment},
 pages = {563--589},
 volume = {29},
 number = {2},
 issn = {08239150},
 journal = {Contemporary Accounting Research},
 doi = {10.1111/j.1911-3846.2011.01114.x       }
}

@article{Muthukrishna.2020,
 abstract = {In this article, we present a tool and a method for measuring the psychological and cultural distance between societies and creating a distance scale with any population as the point of comparison. Because psychological data are dominated by samples drawn from Western, educated, industrialized, rich, and democratic (WEIRD) nations, and overwhelmingly, the United States, we focused on distance from the United States. We also present distance from China, the country with the largest population and second largest economy, which is a common cultural comparison. We applied the fixation index (FST), a meaningful statistic in evolutionary theory, to the World Values Survey of cultural beliefs and behaviors. As the extreme WEIRDness of the literature begins to dissolve, our tool will become more useful for designing, planning, and justifying a wide range of comparative psychological projects. Our code and accompanying online application allow for comparisons between any two countries. Analyses of regional diversity reveal the relative homogeneity of the United States. Cultural distance predicts various psychological outcomes.},
 author = {Muthukrishna, Michael and Bell, Adrian V. and Henrich, Joseph and Curtin, Cameron M. and Gedranovich, Alexander and McInerney, Jason and Thue, Braden},
 year = {2020},
 title = {Beyond Western, Educated, Industrial, Rich, and Democratic (WEIRD) Psychology: Measuring and Mapping Scales of Cultural and Psychological Distance},
 pages = {678--701},
 volume = {31},
 number = {6},
 journal = {Psychological science},
 doi = {10.1177/0956797620916782      },
 file = {http://www.ncbi.nlm.nih.gov/pubmed/32437234},
 file = {https://www.ncbi.nlm.nih.gov/pmc/articles/PMC7357184 }
}

@article{Rosenqvist.2015,
 author = {Rosenqvist, Olof and Skans, Oskar Nordstr{\"o}m},
 year = {2015},
 title = {Confidence enhanced performance? -- The causal effects of success on future performance in professional golf tournaments},
 pages = {281--295},
 volume = {117},
 issn = {01672681},
 journal = {Journal of Economic Behavior {\&} Organization},
 doi = {10.1016/j.        jebo.2015.06.020}
}

@article{Cannon.2005,
 author = {Cannon, Mark D. and Witherspoon, Robert},
 year = {2005},
 title = {Actionable feedback: Unlocking the power of learning and performance improvement},
 pages = {120--134},
 volume = {19},
 number = {2},
 issn = {1558-9080},
 journal = {Academy of Management Perspectives},
 doi = {10.5465/ame.2005.16965107}
}

@article{DeNisi.2000,
 author = {DeNisi, Angelo S. and Kluger, Avraham N.},
 year = {2000},
 title = {Feedback effectiveness: Can 360-degree appraisals be improved?},
 pages = {129--139},
 volume = {14},
 number = {1},
 issn = {1558-9080},
 journal = {Academy of Management Perspectives},
 doi = {10.5465/ame.2000.2909845 }
}

@article{Eggers.2019,
 author = {Eggers, J. P. and Suh, Jung-Hyun},
 year = {2019},
 title = {Experience and Behavior: How Negative Feedback in New Versus Experienced Domains Affects Firm Action and Subsequent Performance},
 pages = {309--334},
 volume = {62},
 number = {2},
 issn = {0001-4273},
 journal = {Academy of Management Journal},
 doi = {10.5465/amj.2017.0046}
}

@article{SullyDeLuque.2000,
 author = {{Sully De Luque}, Mary F. and Sommer, Steven M.},
 year = {2000},
 title = {The Impact of Culture on Feedback-Seeking Behavior: An Integrated Model and Propositions},
 pages = {829--849},
 volume = {25},
 number = {4},
 issn = {0363-7425},
 journal = {Academy of Management Review},
 doi = {10.5465/amr.2000.3707736 }
}

@article{Lee.2021,
  title={The Role of Attribution in Learning from Performance Feedback: Behavioral Perspective on the Choice between Alliances and Acquisitions},
  author={Lee, Jaemin and Lee, Joon Mahn and Kim, Ji-Yub},
  journal={Academy of Management Journal},
  number={(In-press)},
  year={2021}
}

@article{Hattie.2007,
  title={The power of feedback},
  author={Hattie, John and Timperley, Helen},
  journal={Review of educational research},
  volume={77},
  number={1},
  pages={81--112},
  year={2007},
  publisher={Sage Publications Sage CA: Thousand Oaks, CA}
}

@article{Harrison.2015,
  title={An inductive study of feedback interactions over the course of creative projects},
  author={Harrison, Spencer H and Rouse, Elizabeth D},
  journal={Academy of Management Journal},
  volume={58},
  number={2},
  pages={375--404},
  year={2015},
  publisher={Academy of Management Briarcliff Manor, NY}
}

@article{Kim.2020,
  title={Does negative feedback benefit (or harm) recipient creativity? The role of the direction of feedback flow},
  author={Kim, Yeun Joon and Kim, Junha},
  journal={Academy of Management Journal},
  volume={63},
  number={2},
  pages={584--612},
  year={2020},
  publisher={Academy of Management Briarcliff Manor, NY},
  doi={10.5465/amj.2016.1196}
  
}

@article{Rhee.2020,
 author = {Rhee, Mooweon and Alexandra, Valerie and Powell, K. Skylar},
 year = {2020},
 title = {Individualism-collectivism cultural differences in performance feedback theory},
 pages = {343--364},
 volume = {27},
 number = {3},
 issn = {2059-5794},
 journal = {Cross Cultural {\&} Strategic Management},
 doi = {10.1108/CCSM-05-2019-0100 }
}

@article{Berlin.2016,
 author = {Berlin, No{\'e}mi and Dargnies, Marie-Pierre},
 year = {2016},
 title = {Gender differences in reactions to feedback and willingness to compete},
 pages = {320--336},
 volume = {130},
 issn = {01672681},
 journal = {Journal of Economic Behavior {\&} Organization},
 doi = {10.1016/j.     jebo.2016.08.002}
}

@article{Roberts.1994,
 author = {Roberts, Tomi-Ann and Nolen-Hoeksema, Susan},
 year = {1994},
 title = {Gender Comparisons in Responsiveness to Others' Evaluations in Achievement Settings},
 pages = {221--240},
 volume = {18},
 number = {2},
 issn = {0361-6843},
 journal = {Psychology of Women Quarterly},
 doi = {10.1111/j.1471-6402.1994.tb00452.x   }
}

@article{Hofstede.2011,
 author = {Hofstede, Geert},
 year = {2011},
 title = {Dimensionalizing Cultures: The Hofstede Model in Context},
 volume = {2},
 number = {1},
 journal = {Online Readings in Psychology and Culture},
 doi = {10.9707/2307-0919.1014  }
}

	\vfill
	\bibliographystyle{apacite}
	\bibliography{references}
	
	\newpage
	\section*{Appendix}
  
	\subsection*{Descriptive statistics}
	
	\begin{table}[H]
		\caption{Full descriptive statistics} 
		\label{tab:table_desc}
		\begin{tabular}{l r r c r r} 
			\toprule
			& \multicolumn{2}{c}{Diving} & & \multicolumn{2}{c}{Ski jumping} \\
			\cline{2-3} \cline{5-6}
			& Mean & Std. dev. & & Mean & Std. dev. \\ 
			\midrule
			\textit{Treatments:}    ~~~~~~~~~~~~~~~~~~~~~~~~~~~~~~~~~~~             &       &       & &         &       \\
			Positive feedback (deviation positive)          & 0.426   & (0.286) & & 0.316   & (0.262)   \\
			Negative feedback (deviation negative)          & 0.477   & (0.320) & & 0.357   & (0.290)   \\
			Positive feedback$^+$      & 0.314   & (0.297) & & 0.179   & (0.258)   \\		
			Negative feedback$^+$      & 0.363   & (0.328) & & 0.218   & (0.289)   \\
			Future positive feedback   & 0.439   & (0.301) & &         &           \\
			Future negative feedback  & 0.489   & (0.325) & &         &           \\
            \textit{Outcomes:}          &         &         & &         &           \\
			Performance (rem. 3 judges' ratings) & 7.119   & (1.189) & & 17.771  & (0.744)   \\
			Performance (all 5 / 7 judges' ratings)   & 7.110   & (1.182) & &  17.765 & (0.741)   \\
			Score                       & 68.737  & (14.557) & & 118.647 & (16.204) \\
			Distance                    &         &         & & 122.608 &  (11.837) \\
			\textit{Covariates:}        &         &         & &         &           \\
			Difficulty                  &  3.211  & (0.331) & &         &           \\
			Compatriot judge            &  0.248  &         & & 0.457   &           \\
			Home event                  &  0.099  &         & & 0.127   &           \\
			Final                       &  0.291  &         & &         &           \\
			Female                      &   0.450 &         & &         &           \\
			Age                         &  22.429 & (3.789) & & 26.836  & (4.949)   \\
			Current ranking             & 8.490   & (9.655) & &  15.357 & (8.582)   \\
			Start order                 &  9.490  & (11.082) & &         &           \\
			Points behind leader        &  31.491 & (31.011) & & 19.247 & (10.132)  \\
			In range (within 5 pts. to threshold) &  0.264  &     &&    &           \\
			Gate points                  &         &         & &  0.093  & (3.270)   \\
			Wind points                  &         &         & &  -0.291 & (8.225)   \\
			Prev. performance          & 7.270   & (0.958) & & 17.854  & (0.580)   \\
			Prev. SD performance       & 0.130   & (0.151) & & 0.157   & (0.159)   \\
			Prev. wind points            &         &         & &  -1.685 & (8.136)   \\
			Prev. gate points            &         &         & &  -0.163 & (4.386)   \\
			Prev. distance              &         &         & & 123.940 & (11.143)  \\
			Prev. difficulty            & 3.166   & (0.317) & &         &           \\

			\midrule 
			N  &   & 13075 & &   & 4529  \\
			\bottomrule
			\multicolumn{6}{l}{\footnotesize Notes: Mean and standard deviation (in parentheses; for non-binary variables). Some variables }\\
			\multicolumn{6}{l}{\footnotesize ~~~~~~~~~ only observed in one of the data sets. $^+$Alternative definition as defined in the main text.}
			
		\end{tabular}
		
	\end{table}

	\subsection*{Placebo and balancing tests}
	
	\begin{table}[H]
		\centering
		\caption{Placebo treatment regressions} 
		\label{tab:table_placebo_div}
		\begin{tabular}{l r r r r r r r} 
			\toprule
			& \multicolumn{1}{c}{Judges'} & ~~ & \multicolumn{1}{c}{Judges'} & ~~ & \multicolumn{1}{c}{Judges'} & ~~ & \multicolumn{1}{c}{}\\
			& \multicolumn{1}{c}{ratings 3} & ~~ & \multicolumn{1}{c}{ratings 5} & ~~ & \multicolumn{1}{c}{ratings 7} & ~~ & \multicolumn{1}{c}{Score}\\
			\cline{2-2} \cline{4-4} \cline{6-6} \cline{8-8} \\
			Future positive feedback~~~~ & 0.028   & & 0.030   & &  0.027  & & -0.012  \\
			                           & (0.035) & & (0.035) & & (0.035) & & (0.345) \\
			Future negative feedback   & -0.045  & & -0.043  & & -0.041  & & -0.437  \\
			                           & (0.035) & & (0.035) & & (0.035) & & (0.318) \\
			\midrule 
			N                          & 10256   & & 10256   & & 10256   & & 10256   \\ 
			\bottomrule
			\multicolumn{8}{l}{\footnotesize Notes: Linear Regression on the outcome mentioned in the column header. 3, 5, and 7 refer }\\
			\multicolumn{8}{l}{\footnotesize ~~~~~~~~~  to discarding four, two, or none of the extreme judges' ratings. Diving data. Pseudo-}\\
			\multicolumn{8}{l}{\footnotesize ~~~~~~~~~  treatment is the deviation of next (future) jump. Jumps 2--4/5 only. Specifications as }\\
			\multicolumn{8}{l}{\footnotesize ~~~~~~~~~  in column (3) in Table \ref{tab:table_base}. Standard errors are clustered on the individual level. *, **,}\\
			\multicolumn{8}{l}{\footnotesize ~~~~~~~~~   and *** represents statistical significance at the 10 \%, 5 \%, and 1 \% level, respectively. }
		\end{tabular}
	\end{table}

	\begin{table}[H]
		\caption{Balancing Tests} 
		\label{tab:table_bala_test}
		\begin{tabular}{l c c c c c c c c} 
			\toprule
			& \multicolumn{8}{c}{\textbf{Diving}} \\ 
			\cmidrule(){2-9} 
			& \multicolumn{2}{c}{Compatriot judge} & & \multicolumn{2}{c}{Home event} & & \multicolumn{2}{c}{SD prev. perform.}  \\
			\cline{2-3} \cline{5-6} \cline{8-9} 
			\rule{0pt}{3ex} 
			& (1) & (2) & & (3) & (4) & & (5) & (6)  \\ 
			\midrule
			Feedback positive   & -0.000  & & & -0.004  &   & &  0.007  &   \\
			                    & (0.014) & & & (0.008) &   & & (0.005) &   \\
			Feedback negative   & & -0.017  & &   & -0.012  & & & 0.002     \\
			                    & & (0.012) & &   & (0.007) & & &(0.005)    \\
			\midrule 
			& \multicolumn{2}{c}{Difficulty} & & \multicolumn{2}{c}{Final} & & \multicolumn{2}{c}{}  \\
			\cline{2-3} \cline{5-6}
			\rule{0pt}{3ex} 
			& (1) & (2) & & (3) & (4) & &  &   \\ 
			\cline{1-6}
			Feedback positive   & -0.005  & & & -0.018  &   & &  &    \\
			                    & (0.007) & & & (0.013) &   & &  &    \\
			Feedback negative   & & 0.006   & &   & -0.017  & &  &    \\
			                    & & (0.006) & &   & (0.013) & &  &    \\
			\midrule 
			& \multicolumn{8}{c}{\textbf{Ski jumping}} \\ 
			\cmidrule(){2-9} 
			& \multicolumn{2}{c}{Compatriot judge} & & \multicolumn{2}{c}{~~~~~~Home event~~~~~~} & & \multicolumn{2}{c}{Prev. distance}  \\
			\cline{2-3} \cline{5-6} \cline{8-9}  
			\rule{0pt}{3ex} 
			& (1) & (2) & & (3) & (4) & & (5) & (6)  \\ 
			\midrule
			Feedback positive   & -0.045  & & & -0.023  &       & &  0.918  &   \\
			                    & (0.028) & & & (0.018) &       & & (0.696) &   \\
			Feedback negative   & & -0.010  & &   & -0.048***   & & &  0.345    \\
			                    & & (0.022) & &   & (0.017)     & & & (0.674)   \\
			\midrule 
			& \multicolumn{2}{c}{Prev. gate} & & \multicolumn{2}{c}{SD prev. perform.} & & \multicolumn{2}{c}{}  \\
			\cline{2-3} \cline{5-6} 
			\rule{0pt}{3ex} 
			& (1) & (2) & & (3) & (4) & &  &   \\ 
			\cline{1-6}
			Feedback positive   & -0.304  & & &  0.024  &   & &  &    \\
			                    & (0.306) & & & (0.092) &   & &  &    \\
			Feedback negative   & & 0.031   & &   & 0.018  & &  &    \\
			                    & & (0.217) & &   & (0.095) & &  &    \\
			\bottomrule
			\multicolumn{9}{l}{\footnotesize Notes: Linear Regression estimates. Each regression includes athlete fixed-effects. Standard errors}\\
			\multicolumn{9}{l}{\footnotesize ~~~~~~~~~  are clustered on the individual level. *, **, and *** represents statistical significance at the}\\
			\multicolumn{9}{l}{\footnotesize ~~~~~~~~~    10 \%, 5 \%, and 1 \%, respectively.}
		\end{tabular}
		
	\end{table}

	\subsection*{Additional and full results tables}
	
	\begin{table}[H]
		\centering
		\caption{Feedback on performance -- sensitivity to different specifications, ski jumping} 
		\label{tab:table_base_ski}
		\begin{tabular}{l r r r r} 
			\toprule
			& \multicolumn{4}{c}{\textbf{Ski jumping}} \\ 
			\cline{2-5}
			\textbf{Performance} ~~~~~~~~~~~~~~~~~~~  & ~~~~ (1)   ~~~~    &  ~~~~ (2) ~~~~ & ~~~~ (3) ~~~~  & ~~~~ (4) ~~~~  \\ 
			\midrule
			Positive feedback   & 0.201***  & 0.180***  & 0.145***  & 0.107***  \\
			                    & (0.035)   & (0.036)   & (0.034)   & (0.034)   \\
			Negative feedback   & -0.063    &  -0.055   & -0.049    & -0.026    \\
			                    & (0.043)   &  (0.041)  & (0.037)   & (0.041)   \\ 
			Prev. jury assessment  & 0.593***  &  0.465*** & 0.402***  & 0.329***  \\
			                    & (0.027)   &  (0.041)  & (0.043)   & (0.031)   \\
			Prev. wind points   &           &  0.044*** & 0.040***  & 0.036***  \\
			                    &           &   (0.002) &  (0.002)  & (0.002)   \\ 
			Prev. gate points   &           &  0.003    & 0.002     & 0.000     \\
			                    &           &  (0.003)  &  (0.003)  & (0.003)   \\
			Prev. distance      &           & 0.002***  & 0.003***  & 0.002**   \\
			                    &           &  (0.001)  &  (0.001)  &  (0.002)  \\
			Wind points         &           & -0.041*** & -0.038*** & -0.036*** \\
			                    &           &   (0.002) &  (0.002)  & (0.002)   \\ 
			Gate points         &           & -0.020*** & -0.019*** & -0.019*** \\
			                    &           &  (0.003)  &  (0.002)  & (0.003)   \\ 
			Points behind       &           & -0.015*** & -0.016*** & -0.015*** \\
			                    &           &  (0.002)  &  (0.002)  &  (0.002)  \\
			Compatriot judge    &           &  0.021    & 0.016     & 0.024     \\
			                    &           &  (0.020)  & (0.022)   & (0.023)   \\ 
			Home event          &           &  0.013    & 0.028     & 0.041     \\
			                    &           &  (0.032)  & (0.032)   & (0.035)   \\ 
			Start order         &           &  0.002    & -0.003    & -0.005*   \\
			                    &           &  (0.002)  &  (0.002)  &  (0.003)  \\
			SD prev. judges' ratings. &           & -0.019    & -0.001    & -0.017    \\
			                    &           & (0.061)   &  (0.063)  &  (0.065)  \\
			\midrule 
			Athlete Fixed Effect  & &   &  x &    \\ 
			Athlete x Season FE  &  &  &   & x    \\ 
			
			\midrule 
			N  & 4529 & 4529 & 4529 & 4529  \\ 
			\bottomrule
			\multicolumn{5}{l}{\footnotesize Notes: Linear regression. Prev. (= previous) refers to a lagged variable from the previous jump.}\\
			
			\multicolumn{5}{l}{\footnotesize  ~~~~~~~~~   SD = standard deviation. Standard errors are clustered on the individual level. *, **,}\\
			\multicolumn{5}{l}{\footnotesize ~~~~~~~~~   and *** represents statistical significance at the 10 \%, 5 \%, and 1 \% level, respectively.}\\
			
		\end{tabular}
	\end{table}
	
	\begin{table}[H]
		\centering
		\caption{Feedback on performance -- sensitivity to different specifications, diving} 
		\label{tab:table_base_div}
		\begin{tabular}{l r r r r r r r} 
			\toprule
			& \multicolumn{7}{c}{\textbf{Diving}} \\ 
			\cmidrule(l){2-8}
			\textbf{Performance} ~~~~~~~~~~~         & (1)   & ~ & (2)  & ~ & (3)   & ~ & (4)     \\ 
			\midrule
			Positive Feedback       & 0.242*** & & 0.208*** & & 0.115*** & & 0.100***  \\
			                        & (0.036)  & & (0.034)  & & (0.032)  & & (0.035)   \\
			Negative Feedback       & 0.018    & & 0.024    & &  0.001   & & 0.007     \\
			                        & (0.030)  & & (0.030)  & & (0.029)  & & (0.030)   \\ 
			Prev. jury assessment   & 0.430*** & & 0.284*** & & 0.103*** & & 0.073***  \\
			                        & (0.026)  & & (0.022)  & & (0.016)  & & (0.016)   \\
			Prev. difficulty        & 0.794*** & & 0.540*** & & 0.147    & & 0.228**   \\
			                        & (0.079)  & &  (0.087) & &  (0.091) & &  (0.100)  \\ 
			SD prev. judges' ratings &          & & 0.095    & & 0.056    & & 0.029     \\
			                        &          & &  (0.067) & &  (0.067) & &  (0.070)  \\
			Compatriot judge        &          & & -0.015   & & -0.024   & & -0.016    \\
			                        &          & & (0.024)  & &  (0.022) & &  (0.025)  \\ 
			Home event              &          & & 0.129*** & & 0.164*** & & 0.196***  \\
			                        &          & &  (0.038) & &  (0.045) & &  (0.054)  \\ 
			Current ranking         &          & & -0.020*** & & 0.000   & & 0.011***\\
			                        &          & &  (0.002)  & & (0.002) & &  (0.003)  \\
			Start order             &          & & -0.003*** & & -0.006*** & & -0.009*** \\
			                        &          & &  (0.001)  & &  (0.001) & &  (0.001)  \\
			Points behind           &          & & -0.003*** & & -0.000  & & 0.001   \\ 
			                        &          & &  (0.001)  & &  (0.000) & &  (0.001)  \\
			Penalty                 &          & & -0.288    & & -0.362* & & -0.310    \\ 
			                        &          & &  (0.187)  & &  (0.187) & &  (0.200)  \\
			\midrule
			Jump and Event Fixed Effect     &   && x    && x    && x    \\ 
			Athlete Fixed Effect            &   &&      && x    &&      \\ 
			Athlete x Season Fixed Effects  &   &&      &&      && x    \\ 
			
			\midrule 
			N & 13075 & & 13075 & & 13075 & & 13075 \\ 
			\bottomrule
			\multicolumn{8}{l}{\footnotesize Notes: Prev. (= previous) refers to a lagged variable from the previous jump. SD = standard }\\
			\multicolumn{8}{l}{\footnotesize ~~~~~~~~~ deviation. Fixed effects for \textit{Events} are 1m and 3m Springboard and 10m Platform, and the   }\\
			\multicolumn{8}{l}{\footnotesize ~~~~~~~~~ five (female) or six (male) jumps. Standard errors are clustered on the individual level.  }\\
			\multicolumn{8}{l}{\footnotesize ~~~~~~~~~  *, **, and *** represents statistical significance at the 10 \%, 5 \%, and 1 \%, respectively.}
			
		\end{tabular}
	\end{table}

	\begin{table}[H]
		\caption{Robustness Checks} 
		\label{tab:table_robu}
		\begin{tabular}{l r r c r r} 
			\toprule
			& \multicolumn{2}{c}{Ski jumping} & & \multicolumn{2}{c}{Diving} \\
			\cline{2-3} \cline{5-6} 
			& Positive & Negative & & Positive & Negative \\
			& \multicolumn{2}{c}{Feedback} & & \multicolumn{2}{c}{Feedback} \\ 
			\midrule
			Baseline results   ~~~~~~~~~~~~~~~ & 0.121***  & -0.036   & ~~ & 0.115***  & 0.001  \\
			                                         & (0.033)   & (0.044) & ~~ &  (0.032)  & (0.029) \\
			\rule{0pt}{3ex} \\
			Other outcome variable                   & 0.129***  & -0.070*  & ~~ &  0.111*** & 0.005   \\
			(all ratings, incl. discarded)           & (0.035)   & (0.038) & ~~ &   (0.032) & (0.029) \\
			Treatment definition 2                   & 0.119***  & -0.062   & ~~ & 0.109***  & 0.005   \\
			(Discarded vs. last credited)            & (0.033)   & (0.039) & ~~ &  (0.032)  & (0.030) \\
			Treatment definition 3                   &           &         & ~~ & 0.127***  & 0.007   \\
			(Mean discarded vs. mean credited)       &           &         & ~~ &  (0.043)  & (0.039) \\
			\rule{0pt}{3ex} \\
			Without data cleaning                    & 0.124***  &  -0.056  & ~~ &  0.059*   & 0.017   \\
			                                         & (0.044)   & (0.047) & ~~ &  (0.035)  & (0.031) \\ 
			Without dropping failed attempts         & 0.124***  &  -0.042  & ~~ &  0.109*** & 0.031   \\
			                                         & (0.045)   & (0.042) & ~~ &  (0.037)  & (0.033) \\
			\rule{0pt}{3ex} \\                                         
			Only athletes not sharing                & 0.175***  & -0.044   & ~~ &  0.113*** & -0.021  \\
			nationality with a judge                 & (0.040)   & (0.053) & ~~ &   (0.038) & (0.034) \\
			\rule{0pt}{3ex} \\
			Only jumps with no variance              & 0.132***  & -0.044   & ~~ & 0.143***  & 0.056   \\
			in scoring ratings                       & (0.043)   & (0.051) & ~~ &  (0.043)  & (0.038) \\
			\bottomrule
			\multicolumn{6}{l}{\footnotesize Notes: Linear regression. Every line represents two separate regressions, one in each data set. }\\
			\multicolumn{6}{l}{\footnotesize ~~~~~~~~~ Specification as in column (3) in Table \ref{tab:table_base}. Standard errors are clustered on the individual }\\
			\multicolumn{6}{l}{\footnotesize ~~~~~~~~~ level. *, **, and *** represents statistical significance at the 10 \%, 5 \%, and 1 \%, respectively.}
		\end{tabular}
		
	\end{table}

\end{document}